 \documentclass[final,3p,times]{elsarticle}

\usepackage{amssymb}
\usepackage{amsmath}
\usepackage{amsfonts}
\usepackage{multirow}
\usepackage{multicol}
\usepackage{caption}
\usepackage{subfig}
\usepackage{soul}
\usepackage{graphicx}
\usepackage[version=4]{mhchem}
\usepackage{siunitx}
\usepackage{longtable,tabularx}
\DeclareMathOperator\erf{erf}

\journal{Journal of Computational Physics}

\begin{document}

\begin{frontmatter}



\title{Multi-Group Maximum Entropy Method: Modeling Translational Non-Equilibrium}

\author[label1]{Anthony Chang}
\author[label1]{Narendra Singh}
\author[label1]{Marco Panesi\corref{cor1}} 
\ead{mpanesi@illinois.edu} 

\affiliation[label1]{organization={Center for Hypersonics and Entry Systems Studies (CHESS), \\ University of Illinois at Urbana-Champaign},
            addressline={105 S Goodwin Ave},
            city={Champaign},
            postcode={61801},
            state={Illinois},
            country={USA}}

\cortext[cor1]{Corresponding author} 

\begin{abstract}
The most rigorous physical description of non-equilibrium gas dynamics is rooted in the numerical solution of the Boltzmann equation. Yet, the large number of degrees of freedom and the wide range of both spatial and temporal scales render these equations intractable for many relevant problems. Drawing inspiration from model reduction techniques in statistical physics, this study constructs a reduced-order model for the Boltzmann equation, by combining coarse-graining modeling framework with the maximum entropy principle. This is accomplished by projecting the high-dimensional Boltzmann equation onto a carefully chosen lower-dimensional subspace, resulting from the discretization of the velocity space into sub-volumes. Within each sub-volume, the distribution function is reconstructed through the maximum entropy principle, ensuring compliance with the detailed balance. The resulting set of conservation equations comprises mass, momentum, and energy for each sub-volume, allowing for flexibility in the description of the velocity distribution function. This new set of governing equations, while retaining many of the mathematical characteristics of the conventional Navier-Stokes equations far outperforms them in terms of applicability. The proposed methodology is applied to the Bhatnagar, Gross, and Krook (BGK) formulation of the Boltzmann equation. To validate the model's accuracy, we simulate the non-equilibrium relaxation of a gas under spatially uniform conditions and compare it directly with the analytical solution. Additionally, the model is used to analyze the shock structure of a 1-D standing shockwave across an extensive range of Mach numbers. Notably, both the non-equilibrium velocity distribution functions and macroscopic metrics derived from our model align remarkably with the direct solutions of the Boltzmann equation. These results are further validated by comparing them with available experimental data and simulation outcomes from the direct simulation Monte Carlo method, underscoring the robustness and accuracy of the proposed approach.
\end{abstract}




\end{frontmatter}



\section{Introduction}
\label{sec1}

Understanding and modeling non-equilibrium hydrodynamics accurately is vital for several challenging engineering applications, such as atmospheric reentry \cite{gnoffo1999planetary, colonna2019hypersonic}, hypersonic flight \cite{candler2019rate}, and other high-temperature plasma applications. Additionally, this technology holds immense potential in emerging areas, including nanoflows for miniaturized device development \cite{abgrall2008nanofluidic,guo2015nanofluidic}, nanoscale multi-phase flows \cite{lu2019unified,keshavarzi2009energy}, and for application to modeling of thermal protection systems \cite{mohan2023development}.

Non-equilibrium phenomena in gas dynamics arise from the inability of collisional processes to drive the gas toward equilibrium. This is primarily due to the sharp gradients induced by the dynamics of the flow. Under these conditions, the assumption of small deviation from equilibrium, required for the applicability of the Navier-Stokes equations, becomes invalid, necessitating the need to obtain solutions of the more reliable Boltzmann equation \cite{vincenti1965introduction, colonna2022two}. However, given the complexity of the Boltzmann equation, only approximate solutions of the equations have been obtained.

Historically, efforts to model these non-equilibrium flows began about a century ago in gas dynamics. Before the advent of modern computational capabilities, the emphasis was on analytical methods, which, although effective, often required substantial simplifications. Kinetic theory \cite{chapman1990mathematical, ferziger1972mathematical}, led to the derivation of macroscopic conservation equations, with critical contributions from Chapman and Enskog (CE)\cite{graille2009kinetic}. Notable derivations for multicomponent gases came from Nagnibeda and Kustova \cite{nagnibeda2009non}, and Brun \cite{brun2006introduction}. Although rooted in the Boltzmann equations, the application of the CE method remains restricted to near-equilibrium scenarios. Challenges arise when there are pronounced gradients or in conditions of rarefied gas flow, as traditional modeling assumptions are violated.

The quest for a flexible and accurate fluid description applicable to stronger non-equilibrium conditions saw the rise of moment methods \cite{grad1949kinetic, boccelli2023modeling} and extended thermodynamics \cite{levermore1996moment}. Some, like the Grad method \cite{braginskii1965transport}, combine the moment description with a perturbative approach, which can sometimes restrict their flexibility. In contrast, the non-perturbative nature of Levermore's approach \cite{zhdanov2002transport} offers a broader scope of application \cite{capitelli2012fundamental, loffhagen1996two}. These methods generally do not generate proper transport terms unless the corresponding fluxes are treated as additional unknowns. Furthermore, they can also lead to nonphysical microscopic state distribution functions.

In rarefied flow scenarios, probabilistic particle methods like Direct Simulation Monte Carlo (DSMC) \cite{serikov1999particle, munafo2012investigation, pan2021rovibrationally, pan2019vibrational} and particle-in-cell methods \cite{park2001chemical} are often employed. Nevertheless, their computational complexity becomes a limiting factor in transitional and continuum flows, especially when flow collisionality is significant. Although direct Boltzmann solvers are gaining momentum \cite{singh2020consistent, haack2012conservative, morris2011monte}, their utility is still restricted to more straightforward and more academic scenarios. Attempts to hybridize DSMC and direct Boltzmann solver show promise\cite{oblapenko2020velocity}. Deep learning-based closures for more generalized conservation laws perform better than the NSE \cite{macart2022deep}. However, the lack of interpretability and the need for predictive capability requires more rigorous investigation.

Our work aims to address the limitations of existing models by introducing a flexible and adaptive framework that strikes a balance between upholding physical rigor and ensuring computational efficiency. Incorporating cutting-edge algorithms without compromising or violating fundamental principles, the method ensures physical consistency. The central idea in the proposed model is to combine the solution of the coarse-grained dynamics with the partial equilibrium of the underlying microscopic structure\cite{panesi2011electronic, sahai2017adaptive, liu2015general}. The idea of partial equilibrium suggests the application of the maximum entropy principles to the reconstruction of unresolved scales or physics. This particular choice is of paramount importance, as it ensures physical consistency of the model by enforcing the principle of detailed balance and ensuring the positivity and boundness of the distribution function. Furthermore, it provides a way to estimate the accuracy of the coarse grain representation, bypassing the need to compare predictions against the `exact solution', which is often unavailable. The first steps in this direction can be found in the work of Dubroca \cite{dubroca2002prise}.

Our overarching goal is to model translational, thermal, chemical, and radiative non-equilibrium in a coupled and unified fashion. In this paper, our focus is confined to modeling translational non-equilibrium phenomena. Yet, this very methodology has recently been effectively employed for the modeling of thermal, chemical, and radiative non-equilibrium models, commonly referred to as grouping or binning models \cite{panesi2011electronic, liu2010multi, liu2015general}. These models combine the computational efficiency of traditional multi-temperature models with the physical accuracy of collisional radiative models, gathering significant interest in the community.

The paper is structured as follows: Section 2 delves into the theory of coarse graining and the maximum entropy closure. Section 3 elucidates the numerical implementation. Section 4 presents results related to a 0-D relaxation to equilibrium and a standing shock in argon. Section 5 concludes and provides insights into future work.

\section{Physical Model and Mathematical Formulation}
\subsection{Assumptions}
The focus of this work is on translational non-equilibrium effects. Hence, unless explicitly stated, we shall focus on a monatomic gas composed of a single chemical species and ignore internal atomic structure. Gases with internal degrees of freedom have been extensively discussed in other publications \cite{munafo2014spectral}.

\begin{itemize}
\item The gas is dilute and composed of point particles.
\item There are no external forces.
\item The inert particle interactions are binary collisions.
\item The inelastic and reactive collisions, such as dissociation and ionization, are not accounted for.
\end{itemize}

\begin{figure}[ht]
    \centering
    \includegraphics[width=11cm]{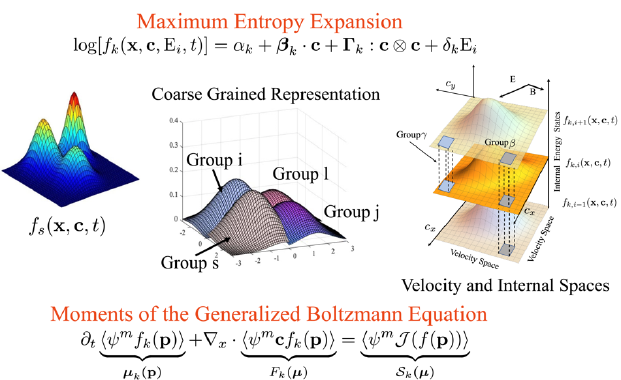}
    \caption{Overview of the maximum entropy-based grouping method. The distribution function is divided into groups in velocity space and reconstructed using the maximum entropy principle.}
    \label{fig:whatisgroup}
\end{figure}

\subsection{Boltzmann Equation}
The evolution of the distribution function, $f({\bf x}, {\bf c}, t)$, \(\boldsymbol{x} = \left(x, y, z\right)\), \(\boldsymbol{c} = \left(c_x, c_y, c_z\right)\), can be obtained by the Boltzmann equation. The distribution function provides the probability of finding particles (molecules, atoms, ions) moving within an infinitesimal velocity range $(d\boldsymbol{c})$  in the neighborhood of velocity $\boldsymbol{c}$ at a given point in physical space. Assuming no body forces and binary collisions, the mathematical form of the Boltzmann equation can be expressed as:

\begin{equation}
    \frac{\partial f}{\partial t} + \boldsymbol{c}\frac{\partial f}{\partial \boldsymbol{x}} = \iint_{\substack{\mathbf{z} \in \mathbb{R}^3 \\ \Omega \in S^2}}\left[f\left(\mathbf{c}^{\prime}\right) f\left(\mathbf{z}^{\prime}\right)-f(\mathbf{c}) f(\mathbf{z})\right] \boldsymbol{g} \frac{\partial \sigma}{\partial \Omega} d \Omega d \mathbf{z}
    \label{boltzmann}
\end{equation}
where the variables \(\boldsymbol{c}\) and \(\boldsymbol{z}\) are the velocities of the two colliding particles. \(\boldsymbol{g} = |\boldsymbol{c} - \boldsymbol{z}|\) is the relative velocity vector, \(\frac{\partial \sigma}{\partial \Omega} d\Omega\) represents the differential cross section integrated over all scattering angles, and \(\,d\boldsymbol{z}\) is the integration over the entire velocity space. Primed variables refer to post-collision velocities and are related to pre-collision velocities by conserving mass, momentum, and energy in an elastic collision \cite{vincenti1965introduction}.

The application of the methodology proposed to the Boltzmann collisional operator is beyond the scope of this work, instead, the model will work with the simplified collision operator developed by Bhatnagar, Gross, and Krook (BGK)\cite{bhatnagar1954model}. The BGK model greatly simplifies the mathematics but retains many qualitative features of the true collision integral, although crucially, the use of the single relaxation time approximation results in a Prandtl number Pr = 1 where experimental data has shown it should be around 2/3 for a monoatomic gas. The right-hand side of Eq.~\ref{boltzmann} is replaced by the simplified collision model:

\begin{equation}
    \frac{\partial f}{\partial t} + \boldsymbol{c}\frac{\partial f}{\partial \boldsymbol{x}} = \nu (f^\text{eq} - f)
    \label{bgk}
\end{equation}
where \(f^\text{eq}\) is the local equilibrium distribution and \(\nu\) is the collision frequency. The collision frequency is computed based on $\nu = P/\mu$, where \(P\) and \(\mu\) are the pressure and viscosity of the gas, respectively.

Any non-equilibrium distribution relaxes to the equilibrium Maxwell-Boltzmann distribution given below, which has been scaled by a number density:
\begin{equation}
f^{\mathrm{eq}}(\mathbf{c})= n\left(\frac{m}{2 \pi k_{\mathrm{B}} T}\right)^{3 / 2} \exp \left(-\frac{m|\mathbf{c}-\mathbf{u}|^2}{2 k_{\mathrm{B}} T}\right)
\label{maxwellian}
\end{equation}
In Eq.~\ref{maxwellian}, \(|\mathbf{c}-\mathbf{u}|\) is the peculiar velocity, \(T\) is the gas temperature, and \(n\) is the number density.  Even with this relatively simple collision operator, Eq.~\ref{bgk} is still computationally intensive to solve for strong non-equilibrium flows.

\section{Multi-Group Maximum Entropy Method}
This section discusses a systematic non-perturbative derivation of the closed systems of new governing equations applicable to the generalized Boltzmann equation. 
The methodology is sketched in Figure \ref{fig:whatisgroup} and draws inspiration from \cite{macdonald2016modeling, liu2015general, sahai2017adaptive}, relies on two steps:
\\
\begin{itemize}
\item	\emph{Local Representation and Reconstruction.} This step relies on a coarse discretization of the velocity space into macroscopic groups and on the reconstruction of the population of each group with the maximum entropy distribution. The reconstruction is obtained by maximizing the entropy within each group, subject to moment constraints chosen to parameterize the distribution. The coefficients are expressed as a function of the moment constraints. Given the generality of the approach, higher-order terms can be included in the expansion.\\

\item \emph{Macroscopic Moment Equations.} Macroscopic governing equations are obtained by taking moments of the Boltzmann equations using the reconstructed local representation. Here, the unknowns are the moment integrals, but other choices are possible (e.g., their quadrature nodal values). The distribution function is reconstructed in terms of the moments and then used to express the flux and source terms in the governing equations (details in Fig.~\ref{fig:whatisgroup}). 
\end{itemize}

\subsubsection{Local Representation and Reconstruction}
In the proposed approach, first, velocity space is divided into several groups, and then the evolution equations for the group density, momentum, and energy are obtained. To derive the macroscopic equations for the groups, an approximation of the probability distribution function within a group is required. Let $\mu^{i}_{k}$ indicate the moments of the distribution function  $f_k ({\bf c}, {\bf x}, t)$ of the $k^{th}$ group:
\begin{equation}
    \boldsymbol{\mu}^{i}_{k} ({\bf x}, t) = \left[\rho_k, \; \rho_k {\bf u}_k, \; \rho_k {e}_k\right]^T = \int\displaylimits_{\boldsymbol{c}_k}^{\boldsymbol{c}_{k+1}} m\left[1, \; {\bf c}, \; {\bf c} \cdot {\bf c} \right]^T f_k ({\bf c})  \, d\boldsymbol{c}, \quad k \in \mathbb{N}
    \label{moment calc}
\end{equation}
The introduction of the above macroscopic quantities as average microscopic quantities with clear physical interpretation, $\left[\rho_k, \; \rho_k {\bf u}_k \; \rho_k {e}_k\right]^T$, which correspond to the mass, momentum and energy of the  $k^{th}$ group. 
Provided $f_k$ is known, one can formulate transport equations for each group by taking moments of the Krook equation (Eq.~\ref{boltzmann}). Therefore, the critical question is to find an approximate functional form for $f_k$.  We apply the maximum entropy-based formulation to a specific group to find an approximation for $f_k$.

{\bf Proposition 1.0.} \emph{To obtain the expression for the reconstruction function, we seek to maximize the entropy, subject to constraints that we choose to describe or characterize the energy state population distribution: } 

\begin{equation}
     \ln(f_k) = \alpha_k + \boldsymbol{\beta}_k \cdot \boldsymbol{c} + \gamma_k \; \boldsymbol{c} \cdot \boldsymbol{c}, \qquad k \in \mathbb{N}
\end{equation}
\emph{where the $\alpha_k$, $\boldsymbol{\beta}_k$ and $\gamma_k$ are functions of the moment constraints $\boldsymbol{\mu^{\nu}_{k}}$.}\\

The expression of the entropy functional of the group $k$ is defined as:
\begin{equation}
    J[f_k] = \int\displaylimits_{c_k}^{c_{k+1}} \left(-f_k \ln f_k + f_k\right)\, d\boldsymbol{c}
    \label{entropy}
\end{equation}
where \(c_k\) and \(c_{k+1}\) are the velocity bounds for the group and \(f_k\) is the group distribution function. The functional form of the group distribution function is obtained to maximize the entropy of the gas in that group and have the first three velocity moments constrained by given values.

Combining the entropy equation and the mass, momentum, and energy constraints, the maximum entropy functional is written as:
\begin{equation}
    J[f_k] = \int\displaylimits_{\boldsymbol{c}_k}^{\boldsymbol{c}_{k+1}} \left(-f_k \ln(f_k) + f_k  + \alpha_k f_k + \boldsymbol{\beta}_k \cdot \boldsymbol{c} f_k + \gamma \boldsymbol{c} \cdot \boldsymbol{c} f_k \right) d\boldsymbol{c} - \alpha_k\mu_k^0 - \boldsymbol{\beta}_k \cdot \boldsymbol{\mu_k^1} - \gamma\mu_k^2
    \label{functional}
\end{equation}
where \(\alpha_k\), \(\boldsymbol{\beta_k}\), and \(\gamma\) are the Lagrange multipliers . From here, take the derivative of the functional with respect to \(f_k\) and set it equal to zero to satisfy the constraints, i.e.
\begin{equation}
    \frac{\partial J}{\partial f_k} = \int\displaylimits_{\boldsymbol{c}_k}^{\boldsymbol{c}_{k+1}} \left(-\ln(f_k) + \alpha_k + \boldsymbol{\beta}_k \cdot \boldsymbol{c} + \gamma \boldsymbol{c} \cdot \boldsymbol{c} \right) d\boldsymbol{c} = 0
\end{equation}
yields, the group distribution function  as:
\begin{equation}
    \ln(f_k) = \alpha_k + \boldsymbol{\beta}_k \cdot \boldsymbol{c} + \gamma \boldsymbol{c}^2 \implies f_k = A_k \exp(-\boldsymbol{\hat{\beta}}_k \left(\boldsymbol{c} - \boldsymbol{w_k}\right)^2), \hspace{0.5 cm} \boldsymbol{c}_k < \boldsymbol{c} < \boldsymbol{c}_{k+1}
    \label{group dist func} 
\end{equation}
Notice that  Eq.~\ref{group dist func} is the equilibrium distribution or Maxwell-Boltzmann distribution. The group distribution function can be expressed in terms of the group distribution parameters, \(A_k\), \(\boldsymbol{\beta}_k\), and \(w_k\).

Physically, the \(\boldsymbol{\beta}_k\) and \(A_k\) terms relate to the inverse temperature of the group, and \(w_k\) relates to the average velocity of the Gaussian interpolant. Each group has its own unique velocity distribution function constrained by the values of the moments with its bounds, and together, the groups form a piece-wise distribution function across the velocity domain. In summary, with this approximation, particles within a group in velocity space are assumed to be in equilibrium with each other and particles in different groups in velocity space can be in non-equilibrium. Instead of evolving a velocity distribution function forward in time, the moments within each group are integrated in time and then the velocity distribution function can be solved for from the macroscopic moments.

\subsubsection{Macroscopic Moment Equations}
In this section, the governing equations are derived. Starting with Eq.~\ref{bgk}, the overall velocity distribution function is replaced by the group distribution function.
\begin{equation}
    \frac{\partial f_k}{\partial t} + \boldsymbol{c}\frac{\partial f_k}{\partial \boldsymbol{x}} = \nu\left(f_k^\text{eq} - f_k\right)
    \label{group bgk}
\end{equation}
By taking the first five velocity moments of the group distribution function, the expressions for the collisional invariants as a function of group distribution parameters can be found.

\begin{equation}
    \boldsymbol{U}_k = \begin{Bmatrix}
        \rho_k \\ \rho_k {\bf u}_k \\ \rho_k e_k
    \end{Bmatrix} = \int\displaylimits_{c^k_x}^{c^{k+1}_x}\int\displaylimits_{c^k_y}^{c^{k+1}_y}
    \int\displaylimits_{c^k_z}^{c^{k+1}_z}
    m
    \begin{pmatrix}
    1\\
    {\bf c} \\
    {\bf c} \cdot {\bf c}
    \end{pmatrix} f_k \,dc_z \,dc_y \, dc_x = \rho  A_k \begin{Bmatrix}
        \mathfrak{I}_{U, 1} \\ \mathfrak{I}_{U, 2} \\ \mathfrak{I}_{U, 3}
    \end{Bmatrix}
     \label{uk}
\end{equation}

\begin{align*}
   & \mathfrak{I}_{U,1} = \mathfrak{I}_i^0\mathfrak{I}_j^0\mathfrak{I}_k^0 \\
   & \mathfrak{I}_{U,2} = \sum_{i=x}^{z}\left(\mathfrak{I}_{i}^1 + w_i\mathfrak{I}_{i}^0\right)\mathfrak{I}_j^0\mathfrak{I}_k^0  \, {\bf e}_i \\
   & \mathfrak{I}_{U,3} = \sum_{i=x}^{z}\left(\mathfrak{I}_i^2 + 2w_i\mathfrak{I}_i^1 + w_i^2\mathfrak{I}_i^0\right)\mathfrak{I}_j^0\mathfrak{I}_k^0 \\
   & i, j, k \in \{x, y, z\} \mid i \neq j \neq k
 \label{uk expr}
\end{align*}

\noindent where \(\mathfrak{I}^n_{i}\) are particular integrals of the following form:

\begin{align}
    \mathfrak{I}^n_{i} = \int_{c_i^{k}}^{c_i^{k+1}} \left(c_i - w_i\right)^n \exp\left[-\beta\left(c_i - w_i\right)^2\right] \, dc_i
\end{align}

\noindent The expressions for the relevant particular integrals used are listed in Appendix A. In Eq.~\ref{uk}, \(\boldsymbol{u}\) is the velocity vector \(\boldsymbol{u} = \left(u, v, w\right)\), with the notation chosen to match general convention for velocity in each Cartesian direction. The flux function can be derived by the following integration:

\begin{equation}
    \boldsymbol{F}_k = \begin{Bmatrix}
            \boldsymbol{F}_1 \\ \Bar{\Bar{F}}_2 \\ \boldsymbol{F}_3
    \end{Bmatrix} = \int\displaylimits_{c^k_x}^{c^{k+1}_x}\int\displaylimits_{c^k_y}^{c^{k+1}_y}
    \int\displaylimits_{c^k_z}^{c^{k+1}_z}
    m
    \begin{pmatrix}
    \bf c\\
     {\bf cc^T} \\
    \left({\bf cc^T}\right) \cdot {\bf c}
    \end{pmatrix} f_k \,dc_y \,dc_z \, dc_x = \rho  A_k \begin{Bmatrix}
        \mathfrak{I}_{F, 1} \\ \mathfrak{I}_{F, 2} \\ \mathfrak{I}_{F, 3}
    \end{Bmatrix} \\ 
    \label{fk eq}
\end{equation}

\begin{align*}
    & \mathfrak{I}_{F,1} = \sum_{i=x}^{z}\left(\mathfrak{I}_{i}^1 + w_i\mathfrak{I}_{i}^0\right)\mathfrak{I}_j^0\mathfrak{I}_k^0 \, {\bf e}_i \\
    & \mathfrak{I}_{F,2} =  \left[\left(\mathfrak{I}_i^2 + 2w_i\mathfrak{I}_i^1 + w_i^2\mathfrak{I}_i^0\right)\mathfrak{I}_j^0\mathfrak{I}_k^0 \delta_{mn} + \left(\mathfrak{I}_i^1 + w_i\mathfrak{I}_i^0\right)\left(\mathfrak{I}_j^1 + w_j\mathfrak{I}_j^0\right)\mathfrak{I}_k^0\left(1 - \delta_{mn}\right)\right] \text{e}_{mn} \\
    & \mathfrak{I}_{F,3} = \sum_{i=x}^{z}\Big[ \left(\mathfrak{I}_i^3 + 3w_i\mathfrak{I}_i^2 + 3w_i^2\mathfrak{I}_i^1 + w_i^3\mathfrak{I}_i^0\right)\mathfrak{I}_j^0\mathfrak{I}_k^0 + \left(\mathfrak{I}_j^2 + 2w_j\mathfrak{I}_j^1 + w_j^2\mathfrak{I}_j^0\right)\left(\mathfrak{I}_i^1 + w_i\mathfrak{I}_i^0\right)\mathfrak{I}_k^0 \, + \\ & \hspace{4.95cm}+\left(\mathfrak{I}_k^2 + 2w_k\mathfrak{I}_k^1 + w_k^2\mathfrak{I}_k^0\right)\left(\mathfrak{I}_i^1 + w_i\mathfrak{I}_i^0\right)\mathfrak{I}_j^0\Big] \, {\bf e}_i \\
    & i, j, k \in \{x, y, z\} \mid i \neq j \neq k ,\quad m, n \in \{1, 2, 3\}
\end{align*}

\noindent where \(m, n\) are the indices of the momentum flux tensor represented by \(\mathfrak{I}_{F,2}\). Finally, replacing \(f_k\) in Eq.~\ref{group bgk} with each of the moments, the governing equations within each group can be written as:
\begin{equation}
    \frac{\partial \rho_k}{\partial t} + \nabla \cdot \boldsymbol{F}_1 = \nu\left(\rho_k^\text{eq} - \rho_k \right)
    \label{gov rho}
\end{equation}
\begin{equation}
    \frac{\partial \rho_k \boldsymbol{u}_k}{\partial t} + \nabla \cdot \bar{\bar{F}}_2 = \nu\left(\rho_k^\text{eq}\boldsymbol{u}_k^\text{eq} - \rho_k \boldsymbol{u}_k \right)
    \label{gov rhou}
\end{equation}
\begin{equation}
        \frac{\partial \rho_k e_k}{\partial t} + \nabla\cdot \boldsymbol{F}_3 = \nu\left(\rho_k^\text{eq}e_k^\text{eq} - \rho_k e_k \right)
    \label{gov rhoe}
\end{equation}
\noindent These equations are qualitatively very similar to the Navier-Stokes equations as they describe the evolution of mass, momentum, and energy in time.

\section{Numerical Implementation}
In this section, the numerical implementation of the governing equations \ref{gov rho} - \ref{gov rhoe} is described. While the equations are derived for a three-dimensional flow, the current work will restrict the grouping to the x-direction which represents zero/one-dimensional flows. This is done to simplify the implementation and validate the maximum entropy model on canonical test cases. The velocity space is kept three-dimensional in all cases and the y and z-velocity is integrated from \((-\infty, \infty)\). Following the same procedure as the previous section, the equations for the moments and the flux function are:
\begin{equation}
   \boldsymbol{U_k} = \begin{bmatrix}
        \rho_k \\ \rho_k u_k \\ \rho_k e_k
    \end{bmatrix} = A_k \begin{bmatrix}
       \frac{\pi}{\beta} \mathfrak{I}^0_x \\
       \frac{\pi}{\beta}\left(\mathfrak{I}^1_{x} + w_x\mathfrak{I}^0_{x}\right) \\
       \frac{\pi}{\beta}\left(\mathfrak{I}^2_{x} + w_x^2 \mathfrak{I}^0_{x} + 2w_x\mathfrak{I}^1_{x} + \frac{\mathfrak{I}^0_{x}}{\beta}\right)
   \end{bmatrix}
   \label{uk eq}
\end{equation}
\begin{equation}
   \boldsymbol{F_k} = \begin{bmatrix}
       F_1 \\ F_2 \\ F_3
   \end{bmatrix} = A_k \begin{bmatrix}
       \frac{\pi}{\beta}\left(\mathfrak{I}^1_{x} + w_x\mathfrak{I}^0_{x}\right) \\
       \frac{\pi}{\beta}\left(\mathfrak{I}^2_{x} + w_x^2 \mathfrak{I}^0_{x} + 2w_x\mathfrak{I}^1_{x}\right) \\
       \frac{\pi}{\beta}\left(\mathfrak{I}^3_{x} + w_x^3 \mathfrak{I}^0_{x} + 3w_x^2 \mathfrak{I}^1_{x} + 3w_x\mathfrak{I}^2_{x}\right) + \frac{\pi(\mathfrak{I}^1_{x} + w_x\mathfrak{I}^0_{x})}{\beta^2}
       \label{fk eq simple}
   \end{bmatrix}
\end{equation}

To solve the Boltzmann equation, it must be integrated in both time and space along with the evaluation of the collision operator. For 1-D flow in the x-direction, Eq.~\ref{gov rhou} simplifies to:
\begin{equation}
    \frac{\partial \rho_k u_k}{\partial t} + \frac{\partial F_2}{\partial x} = \nu\left(\rho_k^\text{eq}u_k^\text{eq} - \rho_k u_k \right)
    \label{rhou 1d}
\end{equation}
The other two governing equations are similarly simplified.

First, the phase space must be discretized and an appropriate method chosen for time-marching and evaluation of the spatial derivative. Next, the moments within each group are calculated from the initial condition. To evaluate the flux function at a time step, the group distribution parameters, \(A_k\), \(\beta_k\), and \(w_k\), must be known which requires solving for these parameters from known group moments. Then, the BGK operator is calculated and then the solution is advanced in time. The phase space discretization, time-marching method, and the solving of group distribution parameters is discussed in the next two sections.

\subsection{Phase space discretization and time-marching method}
The velocity space is discretized using a three-dimensional Cartesian coordinate system and is centered about an origin point. Each direction can have it's own semi-length, denoted by \(L_{v, i}\), where \(i \in \{x, y, z\}\). For simplicity, future test cases will use a cubic velocity grid, setting \(L_{v, x} = L_{v, y} = L_{v, z}\).
The velocity nodes are equally spaced in each direction within the domain given by:
\begin{equation}
    \Delta v = \frac{2L_{v, i}}{N_{v, i}} 
\end{equation}
where \(N_{v,i}\) denotes the number of velocity nodes in the \(i\)-direction. The bounds of the velocity space are denoted by symbols, \(C_{l, i}\) and \(C_{u, i}\) where \(l\) and \(u\) represent the lower and upper bounds respectively.
Since the model is assumed to be one-dimensional in physical space, the solution is sought along a line. The domain size is given by \(L_x\) and nodes are equally spaced along the domain. The bounds are denoted by \(X_{l}\) and \(X_{u}\), with notation similar to the velocity space. The physical space domain is centered at an origin point \(x = 0\). The time domain is discretized into fixed time steps given by \(\Delta t = t_\text{end}/n_T\), where \(n_T\) is the number of time steps and \(t_\text{end}\) is the simulation end time.

The initial velocity distributions are initialized on the discretized velocity grid. Then, the velocity space is further broken down into groups. This process involves deciding group boundaries such that it sufficiently captures all possible distributions throughout the simulation. In this work, the number of groups is denoted by \(N_g\) and is fixed before the simulation. The group boundaries are decided qualitatively based on knowledge of the underlying BGK solution. In future works, different approaches to deciding the number of groups and their location will be explored so that the MGME method will be applicable when the non-grouped distributions are not known beforehand.

An explicit two-stage Runge-Kutta method is employed to march in time and a first-order finite difference method is used to perform the spatial derivative. Explicit time-marching methods are much simpler to implement with lower memory requirements when compared to implicit methods. The Lax-Friedrichs finite difference method is used since it doesn't require knowledge about the flux function derivative or eigenvalues. These two methods were chosen to due to their simplicity and to focus on the implementation of the method itself.

For interior points on the grid, the flux is calculated as:
\begin{equation}
    \Delta \boldsymbol{U}_{k,i} = \boldsymbol{U}_{k, i}^{n+1} - \boldsymbol{U}_{k, i}^{n} = \frac{\Delta t}{\Delta x}\left(\boldsymbol{F}_{k, i - 1/2}^n - \boldsymbol{F}_{k, i + 1/2}^n\right)
\end{equation}
\begin{align*}
    \boldsymbol{F}_{k, i - 1/2}^n = \frac{1}{2}\left(\boldsymbol{F}_{k, i - 1}^n + \boldsymbol{F}_{k, i}^n\right) + \frac{\Delta x}{2\Delta t}\left(\boldsymbol{U}_{k, i-1}^n - \boldsymbol{U}_{k, i}^n\right)
\end{align*}
\begin{align*}
    \boldsymbol{F}_{k, i + 1/2}^n = \frac{1}{2}\left(\boldsymbol{F}_{k, i + 1}^n + \boldsymbol{F}_{k, i}^n\right) + \frac{\Delta x}{2\Delta t}\left(\boldsymbol{U}_{k, i}^n - \boldsymbol{U}_{k, i+1}^n\right)
\end{align*}

\noindent where \(k\) is the group and \(i\) is the spatial index. From here, the flux derivative is found by evaluating:
\begin{align*}
    k_1 &= \Delta \boldsymbol{U}_{k,i} \\
    k_2 &= \boldsymbol{U}_{k,i} + k_1 \\
    \frac{\Delta \boldsymbol{F}_{k,i}}{\Delta x} &= \frac{1}{2}\left(k_1 + k_2\right)
\end{align*}
The time step is calculated according to \cite{mieussens2000discrete}:
\begin{equation}
    \Delta t = \frac{\text{CFL}}{\text{max}\left(\nu\right) + \text{max}\left(\frac{C_{u,i}}{\Delta x}\right)}
\end{equation}
\noindent where CFL is the Courant-Friedrichs-Lewy number. This equation is valid for the Boltzmann equation with the BGK operator \cite{mieussens2000discrete}. The boundary conditions are established by a fixed inflow boundary and an outflow boundary. 

\subsection{Solving of group distribution parameters via non-linear inversion}
In this subsection, the procedure for estimation of group distribution parameters ($\beta, w,$ and $A$) using group-specific moments ($\rho_{k},u_k$ and $e_k$) is discussed.
The group function parameters are necessary to reconstruct the group distribution function and to compute the flux function. First, the system of three equations \ref{uk eq} can be normalized into a system of two equations by dividing the equation for \(\rho_k u_k\) and \(\rho_k e_k\) by the equation for \(\rho_k\).
\begin{equation}
    \begin{bmatrix}
        u_k \\ e_k
    \end{bmatrix} = \begin{bmatrix}
        \frac{\mathfrak{I}^1_x}{\mathfrak{I}^0_x} + w_x \\
        \frac{\mathfrak{I}^2_x}{\mathfrak{I}^0_x} + w_x^2 + 2w_x\frac{\mathfrak{I}^1_x}{\mathfrak{I}^0_x} + \frac{1}{\beta}
        \label{reduced uk}
    \end{bmatrix}
\end{equation}

\noindent Doing so eliminates the \(A_k\) variable from the system of equations and allows for \(\beta_k\) and \(w_k\) to be solved from Eq.~\ref{reduced uk}. Once, \(\beta_k\) and \(w_k\) are known, \(A_k\) can be solved for using any of the original three moment equations. 

\begin{figure}[ht]
    \centering
    \subfloat[First moment normalized by density.]{\includegraphics[width=8.2cm]{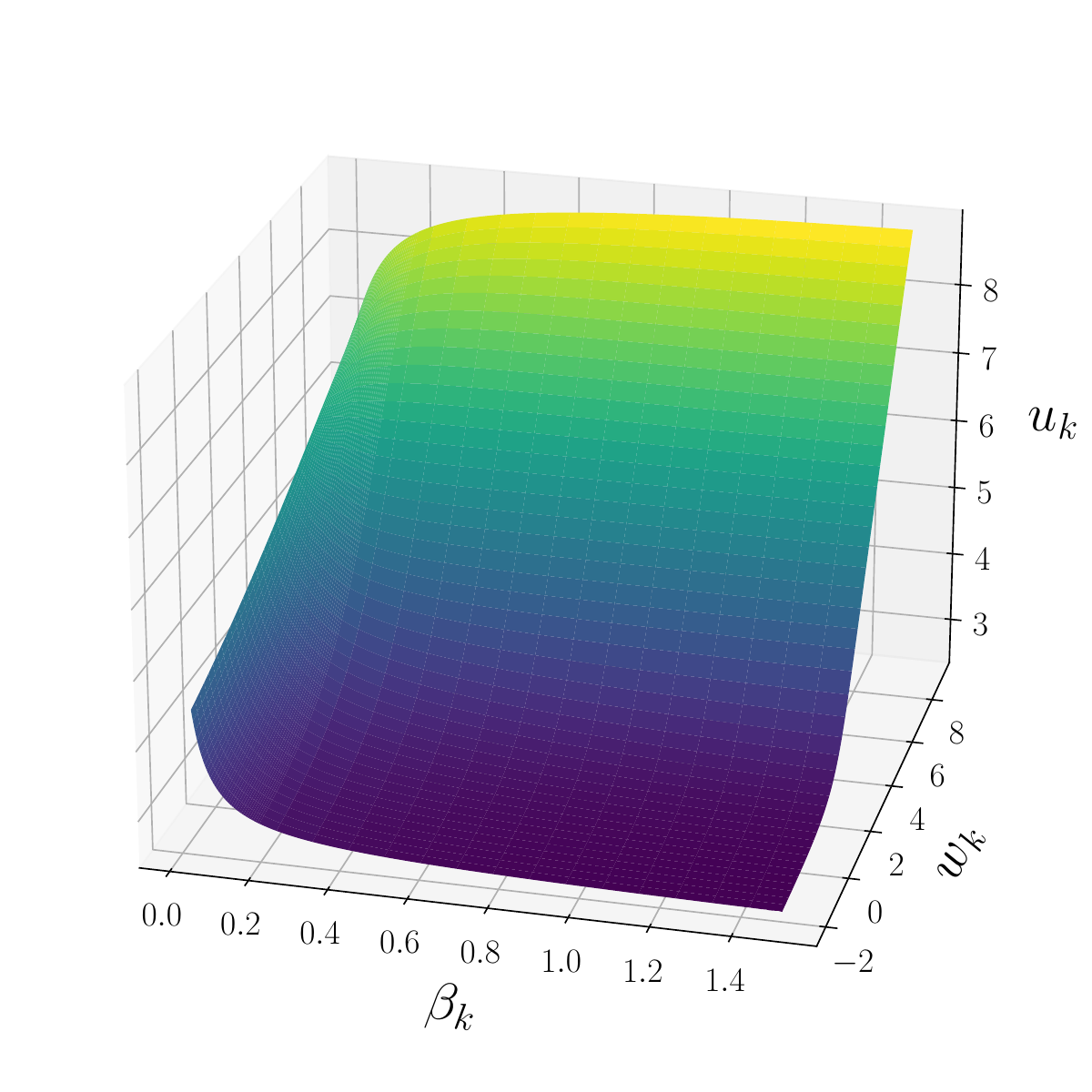}} \hfill
    \subfloat[Second moment normalized by density.]{\includegraphics[width=8.2cm]{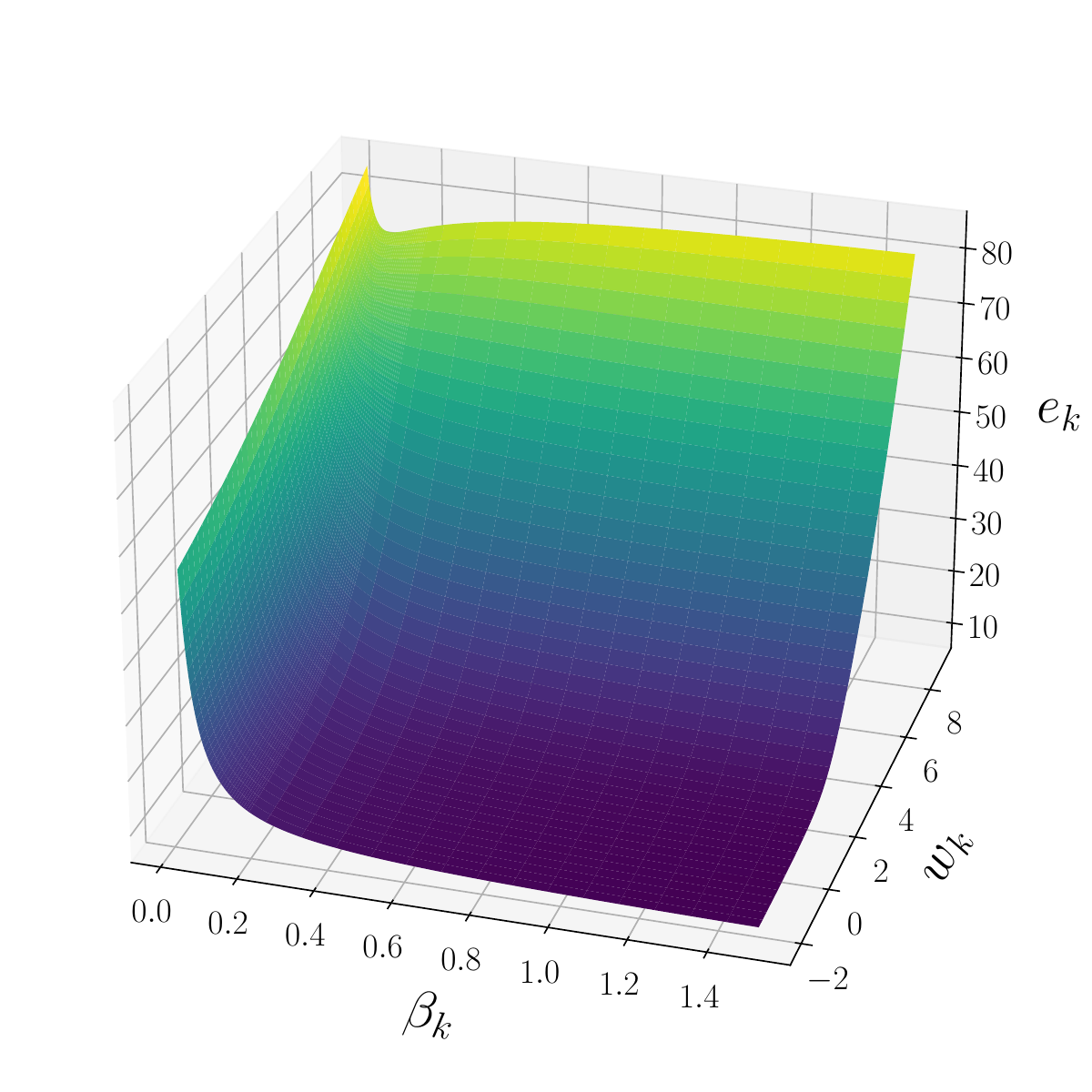}}
    \caption{3D representations of a lookup  table for \(c_i = 2.4\) and \(c_f = 10.0\).}
    \label{fig:inversion plots}
\end{figure}

In Fig.~\ref{fig:inversion plots}, the normalized \(u_k\) and \(e_k\) values are plotted against \(\beta_k\) and \(w_k\). From the plots, the surfaces between the group distribution parameters and the group moment values are smooth but not monotonic. Despite their complex analytic form, the smoothness and boundedness of Eq.~\ref{reduced uk} allows for solutions to be easily found.

In the current work, a combination of a lookup table and a nonlinear equation solver is employed. For the first time step, solutions to the system of equations are sought first using the lookup table and then fed into the nonlinear solver as initial guesses. For future iterations, the previous output of the nonlinear solver is used as the initial guess as the group distribution parameters are expected to evolve smoothly in time. The nonlinear solver uses the \textit{fsolve} function from SciPy which is a wrapper around the MINPACK routine \textit{hybrd}, a modified Powell's dog leg method. The lookup table inversion method to compute the group parameters can be summarized in the following steps:
\begin{enumerate}
    \item Given a normalized moment in a group, for example, \(u_k\),  choose a \(\beta_k\) value in the \(u_k\) table. This returns an array of \(u_k\) values for varying \(w_k\) and fixed \(\beta_k\). Interpolate a \(w_k\) value if \(u_k\) exists for the \(\beta_k\) value chosen. This step is repeated for all \(\beta_k\) values and is done for \(e_k\) as well. The result is two curves, with each curve representing all possible solutions at that specific moment.
    \item Iterate through each \(w_k\) value and interpolate a \(\beta_k\) value similar to the previous step.
    \item Curves of \(\beta_k\) vs. \(w_k\) and vice versa are created. The intersection of the curves represents the solution of the system of equations.
    \item Using the known group parameters \(\beta_k\) and \(w_k\), \(A_k\) can be solved for via Eq.~\ref{uk eq}.
\end{enumerate}
Once the group parameters for each group are obtained, the flux can be calculated using Eq.~\ref{fk eq}. This is followed by the computation of the collision operator for each group. Adding together the contribution from the collision operator and the flux operator, the group moments are updated to the next time step, completing one time step of the algorithm. This process is repeated until a specified steady state is reached.

\subsection{Phase space discretization and time-marching method}
The velocity space is discretized using a three-dimensional Cartesian coordinate system and is centered about an origin point. Each direction can have it's own semi-length, denoted by \(L_{v, i}\), where \(i \in \{x, y, z\}\). For simplicity, future test cases will use a cubic velocity grid, setting \(L_{v, x} = L_{v, y} = L_{v, z}\).
The velocity nodes are equally spaced in each direction within the domain given by:
\begin{equation}
    \Delta v = \frac{2L_{v, i}}{N_{v, i}} 
\end{equation}
where \(N_{v,i}\) denotes the number of velocity nodes in the \(i\)-direction. The bounds of the velocity space are denoted by symbols, \(C_{l, i}\) and \(C_{u, i}\) where \(l\) and \(u\) represent the lower and upper bounds respectively.
Since the model is assumed to be one-dimensional in physical space, the solution is sought along a line. The domain size is given by \(L_x\) and nodes are equally spaced along the domain. The bounds are denoted by \(X_{l}\) and \(X_{u}\), with notation similar to the velocity space. The physical space domain is centered at an origin point \(x = 0\). The time domain is discretized into fixed time steps given by \(\Delta t = t_\text{end}/n_T\), where \(n_T\) is the number of time steps and \(t_\text{end}\) is the simulation end time.

The initial velocity distributions are initialized on the discretized velocity grid. Then, the velocity space is further broken down into groups. This process involves deciding group boundaries such that it sufficiently captures all possible distributions throughout the simulation. In this work, the number of groups is denoted by \(N_g\) and is fixed before the simulation. The group boundaries are decided qualitatively based on knowledge of the underlying BGK solution. In future works, different approaches to deciding the number of groups and their location will be explored so that the MGME method will be applicable when the non-grouped distributions are not known beforehand.

An explicit two-stage Runge-Kutta method is employed to march in time and a first-order finite difference method is used to perform the spatial derivative. Explicit time-marching methods are much simpler to implement with lower memory requirements when compared to implicit methods. The Lax-Friedrichs finite difference method is used since it doesn't require knowledge about the flux function derivative or eigenvalues. These two methods were chosen to due to their simplicity and to focus on the implementation of the method itself.

For interior points on the grid, the flux is calculated as:
\begin{equation}
    \Delta \boldsymbol{U}_{k,i} = \boldsymbol{U}_{k, i}^{n+1} - \boldsymbol{U}_{k, i}^{n} = \frac{\Delta t}{\Delta x}\left(\boldsymbol{F}_{k, i - 1/2}^n - \boldsymbol{F}_{k, i + 1/2}^n\right)
\end{equation}
\begin{align*}
    \boldsymbol{F}_{k, i - 1/2}^n = \frac{1}{2}\left(\boldsymbol{F}_{k, i - 1}^n + \boldsymbol{F}_{k, i}^n\right) + \frac{\Delta x}{2\Delta t}\left(\boldsymbol{U}_{k, i-1}^n - \boldsymbol{U}_{k, i}^n\right)
\end{align*}
\begin{align*}
    \boldsymbol{F}_{k, i + 1/2}^n = \frac{1}{2}\left(\boldsymbol{F}_{k, i + 1}^n + \boldsymbol{F}_{k, i}^n\right) + \frac{\Delta x}{2\Delta t}\left(\boldsymbol{U}_{k, i}^n - \boldsymbol{U}_{k, i+1}^n\right)
\end{align*}

\noindent where \(k\) is the group and \(i\) is the spatial index. The boundaries are prescribed with a fixed inlet boundary condition and an outlet boundary. At the outlet point, the value of the distribution is set to the same as the closest upstream grid point.

The time step is calculated according to \cite{mieussens2000discrete}:
\begin{equation}
    \Delta t = \frac{\text{CFL}}{\text{max}\left(\nu\right) + \text{max}\left(\frac{C_{u,i}}{\Delta x}\right)}
\end{equation}
\noindent where CFL is the Courant-Friedrichs-Lewy number. This equation is valid for the Boltzmann equation with the BGK operator \cite{mieussens2000discrete}. The boundary conditions are established by a fixed inflow boundary and an outflow boundary. 

\section{Results and Discussion}
In this section, we apply the MGME model to analyze non-equilibrium phenomena in shock-heated flow over a wide range of Mach numbers. Extensively studied by Alsmeyer \cite{alsmeyer1976density}, the shock structure analysis serves as a benchmark for evaluating non-equilibrium models. Before investigating the shock problem, we examine a spatially homogeneous, adiabatic relaxation to equilibrium of a monatomic gas starting from strong non-equilibrium conditions. This preliminary test verifies and validates the proposed model's performance in capturing non-equilibrium dynamics.

\subsection{Adiabatic Homogeneous Relaxation}
A spatially homogeneous adiabatic relaxation of the gas is simulated using the MGME model and the BGK-Boltzmann equation. This simple test case serves to verify the accuracy of the coarse-grained model and assess the robustness of the inversion process. The evolution of the microscopic distribution function, the entropy, density, and temperature is bench-marked against the solutions obtained using the BGK-Boltzmann equation.
\subsubsection{Initial state conditions}
For the considered spatially homogeneous relaxation, a symmetrical non-Maxwell-Boltzmann bimodal distribution is initialized and the gas is allowed to relax to equilibrium. The expression for the bimodal distribution, centered around \(c_x = 0\), has the following form:
\begin{equation}
    f(\boldsymbol{c}) =\frac{n}{2}\left(\frac{b}{\pi}\right)^{1.5} \left[\exp(-b(c_x - v)^2) + \exp(-b(c_x + v)^2)\right] \exp(-b(c_y^2 + c_z^2))
    \label{bimodal eq}
\end{equation}
where \(n\) is the number density, and \(b, v\) are constants that determine the exact shape of the distribution. Physically, the above initial distribution corresponds to a gas obtained after mixing gases (in equal amounts) at different equilibrium states, corresponding to identical temperature ($1/(b k_B)$) but different average $x$ velocities ($v$ and the other $-v$). From this bimodal distribution, the initial conditions for the reduced MGME equations ($\rho_k, \rho_k u_k,$ and $\rho_k e_k$) can be obtained by projecting the distribution into groups. Microscopically, the gas relaxes to an equilibrium state characterized by the following Maxwell-Boltzmann distribution:
\begin{equation}
    f^\text{eq}(\boldsymbol{c}) = n\left(\frac{b^\text{eq}}{\pi}\right)^{1.5} \exp(-b^\text{eq}(c_x^2 + c_y^2 + c_z^2)) \  \text{with} \  \frac{3}{2 b^\text{eq}} = \frac{3}{2b} + v^2
\end{equation}
In this test case, \(b = 3\) and \(v = 1\) which results in \(b^\text{eq} = 1\). Since the equilibrium distribution is known a priori, the moment values that the bimodal distribution will relax to can be computed by taking the moments of the equilibrium distribution in each velocity group. The collision frequency is fixed at 0.1 for this test case. 

To assess the accuracy of the solution using various groups, the velocity space domain is divided into two groups and twelve groups, as shown in Fig.~\ref{fig:group 0d}. The dotted lines on the x-axis denote the boundaries of each group. A total of 121 velocity nodes are used with a total domain length in each direction of 6 centered about the zero origin point, \(C_{l,i} = -3, C_{u,i} = 3\). The simulation is run up to a non-dimensional time of 50 for the gas to reach equilibrium. Since there is no space dimension, Eq.~\ref{rhou 1d} reduces to:
\begin{equation}
    \frac{d \left(\rho_k u_k\right)}{d t} = \nu\left(\rho_k^\text{eq}u_k^\text{eq} - \rho_k u_k \right)
    \label{rhou 0d}
\end{equation}
A similar evolution equation for the mass and energy is obtained. It is important to note that, due to the BGK model, the moments in each group evolve independently due to the lack of coupling between groups that appears with the flux function in Eq.~\ref{rhou 1d}. This results in a set of ODE's for the evolution of density, velocity, and temperature in each group.

\subsubsection{Comparisons between MGME method and BGK model}
This section starts with an analysis of the dynamics of the distribution function and later focuses on the macroscopic thermodynamic quantities, such as entropy, density, and temperature. The goal is to assess the validity of the MGME model by comparing it with the BGK model.

\begin{figure}[ht]
    \centering
    \subfloat[2 group solution.]{\includegraphics[width=8.43cm]{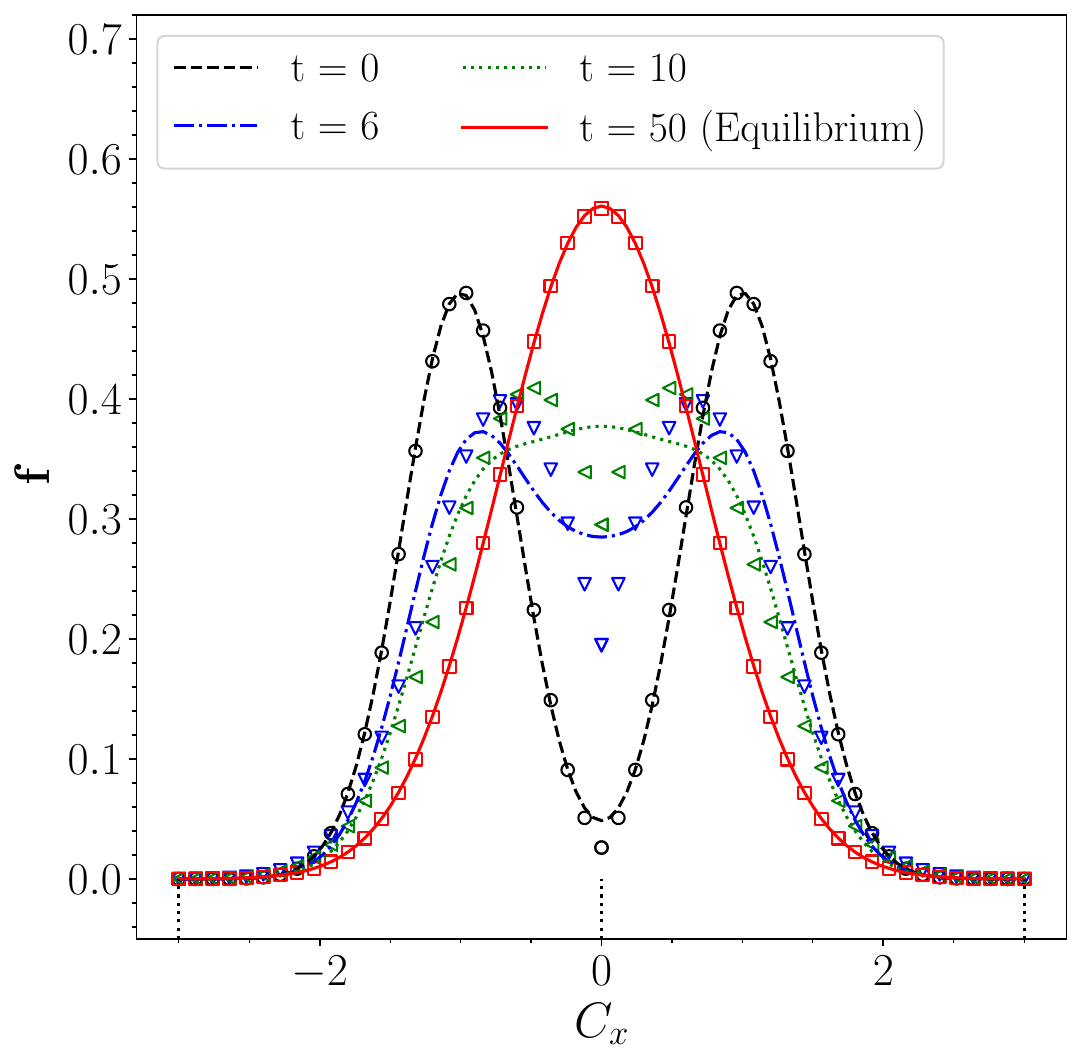}\label{fig:2_group_0d}}
    \subfloat[12 group solution.]{\includegraphics[width=8.03cm]{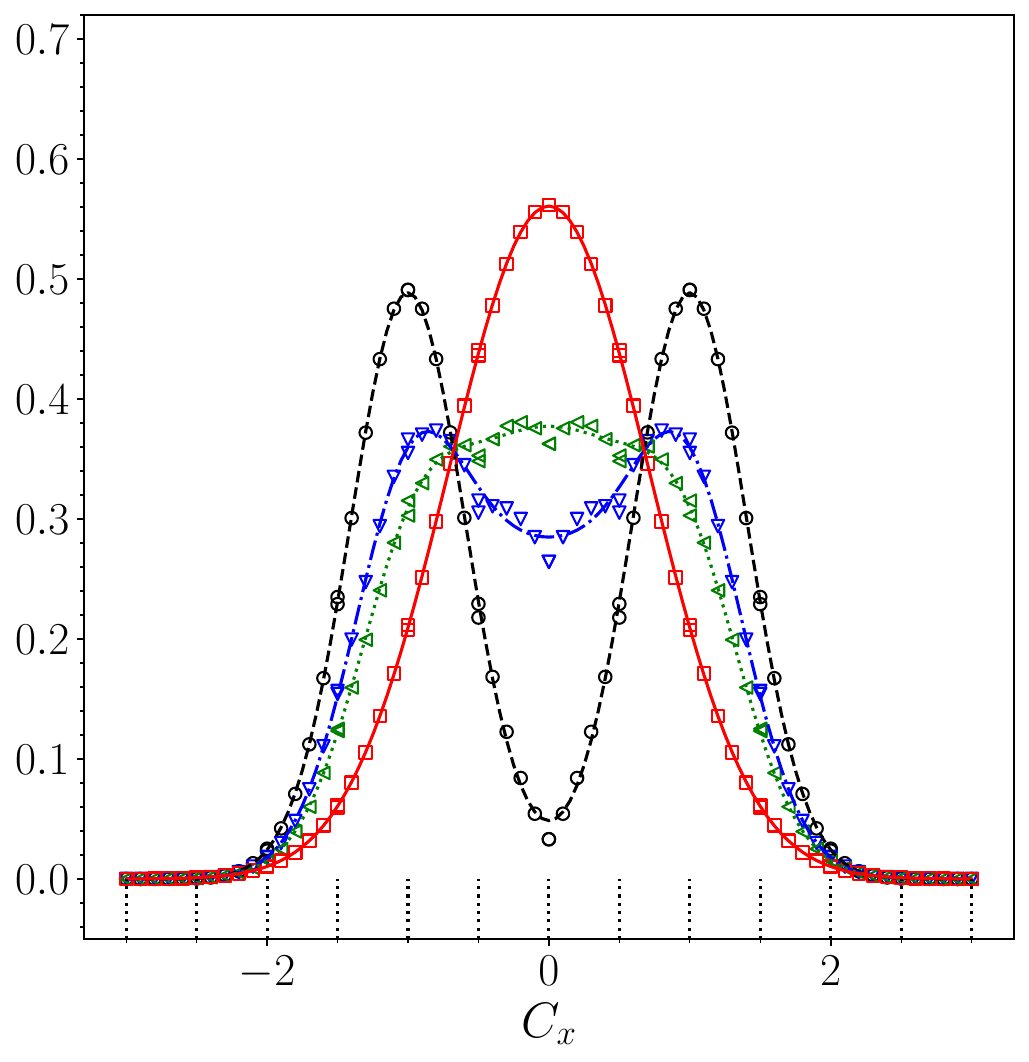}}
    \caption{Evolution of adiabatic homogeneous gas. Solutions obtained using MGME method and BGK-Boltzmann equation denoted by symbols and lines respectively. Group boundaries denoted by dotted vertical lines at the bottom of each graph.}
    \label{fig:group 0d}
\end{figure}

The time evolution of the distribution functions during relaxation, as calculated using both the BGK model and the MGME model, is shown in Fig.~\ref{fig:group 0d}. In this test case, the distribution function represented by the twelve groups in the MGME method shows good agreement with the BGK results. In the same figure, the results obtained with a two-group model give an idea of the impact of the number of groups on accuracy. With only two groups, the initial condition is well represented due to the bimodal shape and choice of groups. However, it struggles for the intermediate distributions due to the fixed shape of the group reconstruction function. It is expected that as the degree of non-equilibrium increases, the more groups are necessary to reconstruct the original distribution function. At equilibrium, the original distribution function would be accurately captured by one group. The two and twelve group solution shows that addition of more groups has no effect on the exact reconstruction of the equilibrium distribution function. 

\begin{figure}[ht]
    \centering
    \subfloat[Entropy]{\includegraphics[width=5.47cm]{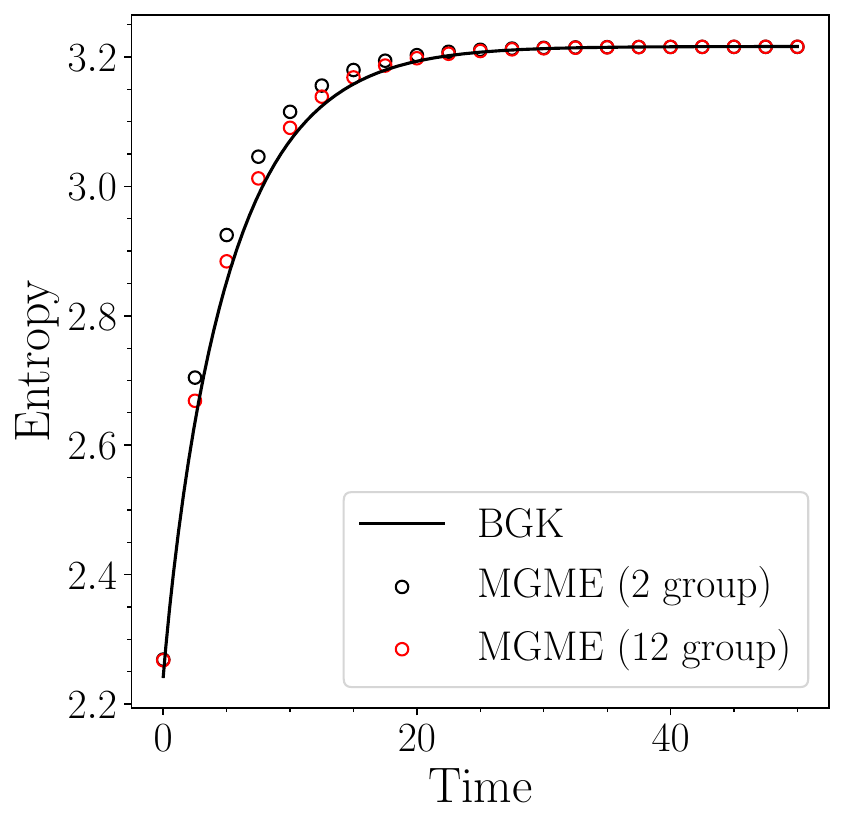}\label{fig:0d total entropy}} 
    \subfloat[Group Entropy]{\includegraphics[width=5.47cm]{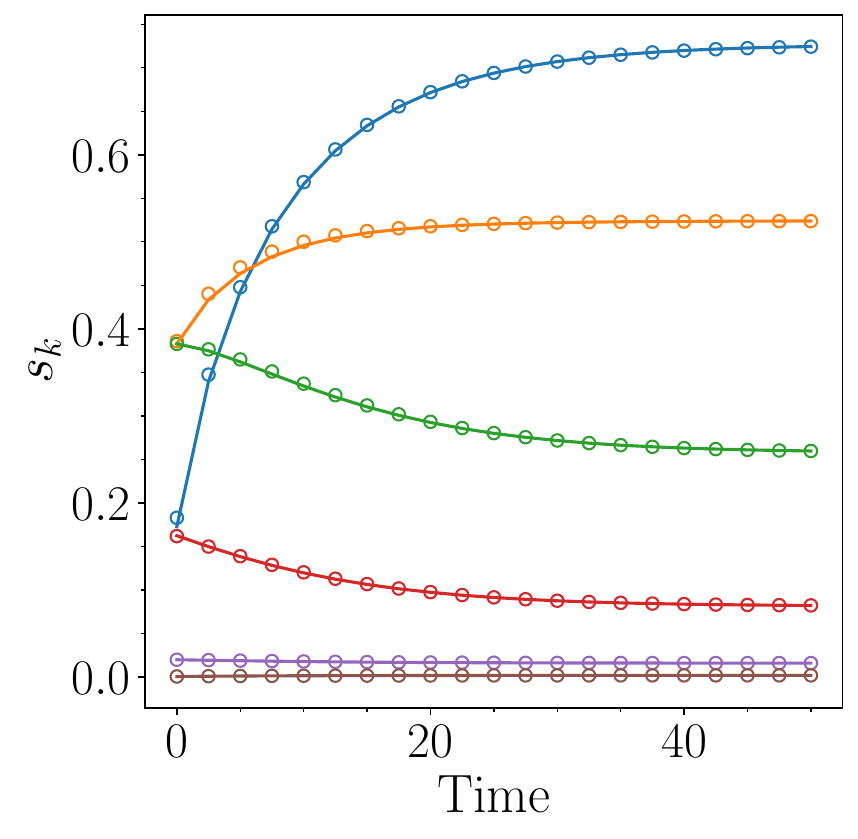}\label{fig:0d group entropy}}
    \subfloat[Group Density]{\includegraphics[width=5.47cm]{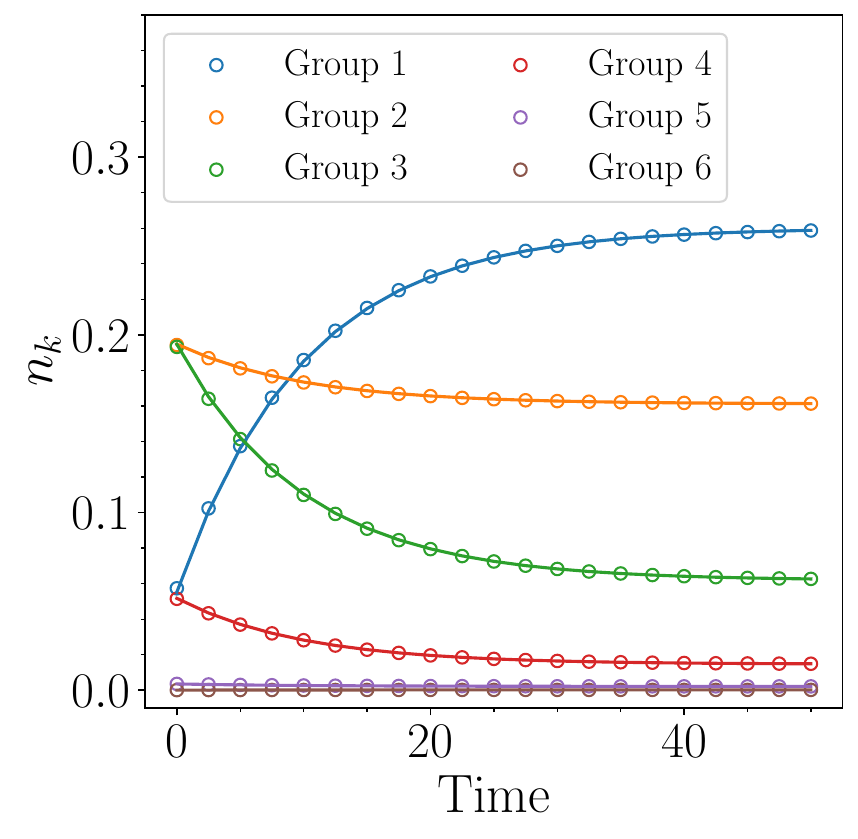}\label{fig:0d group density}} \hfill
    \caption{Time evolution of total and group entropy and group density. In the group variables, solutions obtained by the MGME method are compared to the BGK-Boltzmann solution, denoted by solid lines.}
    \label{fig:0d entropy temp}
\end{figure}

Fig.~\ref{fig:0d entropy temp} shows the time evolution of the total entropy, group density, and group entropy of the gas. For the group parameters, the results of the six positive x-velocity groups in the twelve group solution are shown. Since the evolution of the groups are decoupled, the solution is symmetrical about the zero velocity point. The analysis of the group-specific properties demonstrates excellent agreement of the MGME model with the Boltzmann solution. The total entropy evolution, shown in Fig.~\ref{fig:0d total entropy}, is shown comparing the two grouping solutions along with the Boltzmann solution. As the distribution relaxes to equilibrium, the entropy increases as predicted by the model. It is important to note that the increase in total entropy is enforced by model construction, even though the entropy in a specific group can decrease. Since the entropy is a function of the distribution function values, any differences in the MGME reconstruction would show as a deviation from the Boltzmann solution. This can be seen in the two group solution which deviates from the solid line much more compared to the twelve group solution.

In Fig.~\ref{fig:0d group entropy}, the group entropy for the six groups in the positive x-velocity domain is shown. The group entropy is defined as:

\begin{equation}
    s_k = \int\displaylimits_{c_k}^{c_{k+1}}\int\displaylimits_{-\infty}^\infty\int\displaylimits_{-\infty}^\infty -f_k\ln{\left(f_k\right)} \,dc_y \,dc_z \, dc_x
    \label{group entropy}
\end{equation}

As the system tends towards equilibrium, high-energy molecules transfer their energy to low-energy molecules, resulting in an increase in entropy of the first and second group near the center of the equilibrium distribution and a decrease in entropy in the third and fourth group. Looking at the group entropy, slight deviation is noted for group 2 and 3, suggesting that further grouping in this region of velocity space will result in a more accurate solution. Although not explored in this study, the time evolution of entropy can help determine the optimal number and bounds of velocity groups. The transfer of high-energy molecules to low-energy molecules is also reflected in the time evolution of group density as shown in Fig.~\ref{fig:0d group density}. The group density near the zero velocity increases while group densities outside of this region decreases. There is excellent agreement with the Boltzmann solution since the MGME method enforces conservation. 

It is worth noting, that the conventional Navier-Stokes equations, and more sophisticated models such as higher-order and thirteen-moment equations fail to describe the details of the dynamics discussed in this problem as the macroscopic moments remain unchanged over time. Therefore, the proposed method, which accurately reproduces these non-equilibrium dynamics with significant reductions in the order of velocity space, demonstrates significant advantages in simulating unsteady, non-equilibrium hydrodynamics, where macroscopic quantities are sensitive to non-equilibrium distributions.

\subsection{Argon Shock Structure Analysis}
In this subsection, simulation results for argon flow across a standing shock are presented using the MGME method. The results are compared with the BGK model to evaluate the accuracy of the reduced-order model. Additionally, experimental measurements and available DSMC simulation data are used as reference solutions to assess the strengths and limitations of the MGME method.

\subsubsection{Input conditions and related details of physical parameters}
For the simulations, a calorically perfect argon gas is assumed, and post-shock values for all cases are calculated using the Rankine-Hugoniot jump relations. The input parameters are selected to match the conditions of Alsmeyer \cite{alsmeyer1976density}, and are summarized in Table~\ref{tab:flow parameter}. Three different pre-shock Mach numbers—3, 6, and 10—are considered to study the shock structure under varying flow conditions.

\begin{table}[ht]
    \centering
    \begin{tabular}{c|c|c|c|c|c|c|c|c}
    \hline  
       Parameters &   \(\gamma\) & \(T_\infty\) [K]  &  \(P_\infty\) [Pa]  & \(R\) [J/kg-K] &\(d\) [m]   &  \(T_0\) &\(\omega\) & \(M\) 
        \\
           \hline
           Value & \(\frac{5}{3}\) &  300 & 6.6667 & 208.13 & \num{3.974e-10} & 273.15 & 0.72 & (3, 6, and 10) \\
    \end{tabular}
    \caption{Flow field and viscosity model parameters. }
    \label{tab:flow parameter}
\end{table}

The BGK model requires the calculation of the collision frequency, which is a function of the local pressure and temperature and is computed by a power law model, given by:

\begin{equation}
    \mu = \mu_0\left(\frac{T}{T_0}\right)^\omega \hspace{1.0 in}  \mu_0 = \frac{15 \sqrt{2 \pi m_r k_b T_0}}{2 (5 - 2\omega)(7 - 2\omega)\sigma}
    \label{viscosity}
\end{equation}

where \(m_r\) is the reduced mass and \(\sigma\) is the collision cross section. The reference temperature, \(T_0\), and \(\omega\) are listed in Tab.~\ref{tab:flow parameter}. A hard sphere collision cross-section value of \(\omega\) = 0.7 has been shown to provide good agreement between DSMC results and experimental data \cite{bird1970direct,alsmeyer1976density, reese1995second,lumpkin1992accuracy} for the pre-shock conditions of \(T_1 = 300\) K, \(P_1 = 6.6667\) Pa and \(M_1 = 9\) (see Ref.~\cite{boyd2017nonequilibrium}). In Ref.~\cite{macrossan2003viscosity}, \(\omega = 0.72\) is used as the exponent for DSMC to calculate viscosity for argon at temperatures >2000 K. In Ref.~\cite{alsmeyer1976density}, the experimental results are compared with DSMC simulations using an \(\omega\) value of 0.72. This value reproduces the viscosity measurements of Guevara et al. \cite{guevara1969high} and the viscosity calculations of Amdur et al. \cite{amdur1958properties} from molecular beam experiments. The quantity \(\omega\) depends on $\alpha$ as $ \omega \equiv 2/\alpha + 1/2$, where \(\alpha\) is the exponent relating distance and inter-atomic force, \(F \propto 1/r^\alpha\), necessitating that $\omega > 0.5$ for physical consistency.

From Tab.~\ref{tab:flow parameter}, the model uses \(\omega = 0.72\). This value was primarily chosen by considering the literature and to reach the same freestream mean free path as in Alsmeyer \cite{alsmeyer1976density}.

\subsubsection{Simulation parameters and details}
The non-dimensional velocity grid, space grid, and flow parameters for the standing shock are presented in Tab.~\ref{tab:grid parameter}. The number of groups was fixed at three to compare accuracy across all Mach numbers analyzed.

Pre-shock and post-shock distributions are initialized with Maxwell-Boltzmann distributions with temperatures and densities calculated using the Rankine-Huginot jump conditions with a cosine ramp function applied to create a smooth transition between pre-shock and post-shock distributions.
\begin{table}[ht]
    \centering
    \begin{tabular}{c|c|c|c}
    \hline
    Parameters & Mach 3 & Mach 6 & Mach 10 \\
    \hline
        \(N_g\) & 3 & 3 & 3 \\
        \(N_v\) & 101 & 121 & 131 \\
        \(L_v\) & 14 & 24 & 40 \\
        \(N_x\) & 431 & 481 & 673 \\
        \(L_x\) & 32 & 36 & 50
    \end{tabular}
    \caption{Velocity and physical space grid parameters. \(N_g\) is the number of groups, \(N_v\) is the number of points in velocity space in each direction, \(N_x\) is the number of points in space. }
    \label{tab:grid parameter}
\end{table}

Many grid points in space are used to ensure convergence due to the use of a first-order spatial integration method. All values are non-dimensionalized by a reference value. The density is non-dimensionalized using the pre-shock density value. The velocity and temperature are non-dimensionalized using the following reference values:
\begin{align}
    c_{\text{ref}} = \sqrt{\frac{2  k_BT_\infty}{m_\text{ref}}} \hspace{1.0in}
    T_\text{ref} = \frac{m_\text{ref}}{2 k_B} c_\text{ref}^2 = T_\infty
\end{align}
where \(m_\text{ref}\) is the reference mass, in this case argon, and \(T_\infty\) is the pre-shock temperature. Using these reference values, the reference values for heat flux and shear stress are defined as:
\begin{align}
    \tau_{\text{ref}} = 2 n_\text{ref} k_B T_\text{ref} \hspace{1.0in} q_\text{ref} = \frac{m_\text{ref}}{n_\text{ref}}c_\text{ref}^3
\end{align}
The spatial domain is non-dimensionalized by the reference mean free path given by:
\begin{equation}
    \lambda_\infty = \frac{16}{5}\left(\frac{\gamma}{2\pi}\right)^{1/2} \frac{\mu_\infty}{\rho_\infty a_\infty}
    \label{ref mean free path}
\end{equation}
\noindent where \(\mu_\infty\), \(\rho_\infty\), and \(a_\infty\) are the pre-shock viscosity, density, and speed of sound respectively. The speed of sound is calculated using \(\sqrt{\gamma R T_\infty}\) and the viscosity is calculated using Eq.~\ref{viscosity}. For the free stream variables considered, the reference mean free path is 1.096 mm.

\subsubsection{Velocity distribution functions}
\begin{figure}[ht]
    \centering
    \subfloat[Mach 10, center of shock.]{\includegraphics[width=5.55cm]{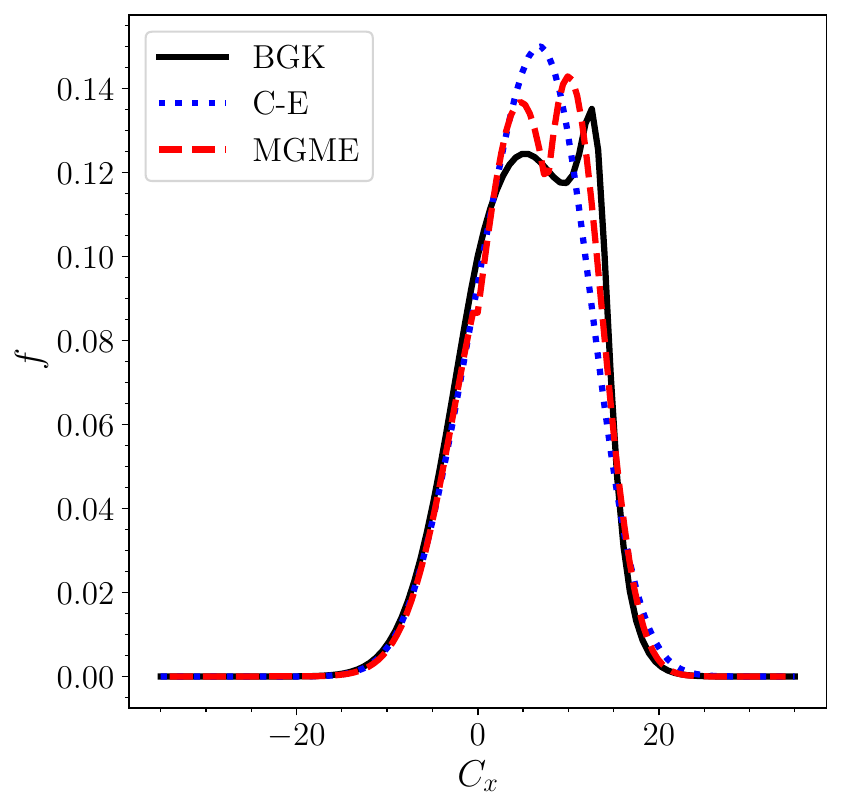}}
    \subfloat[Mach 10, one \(x/\lambda_\infty\) upstream of center.]{\includegraphics[width=5.27cm]{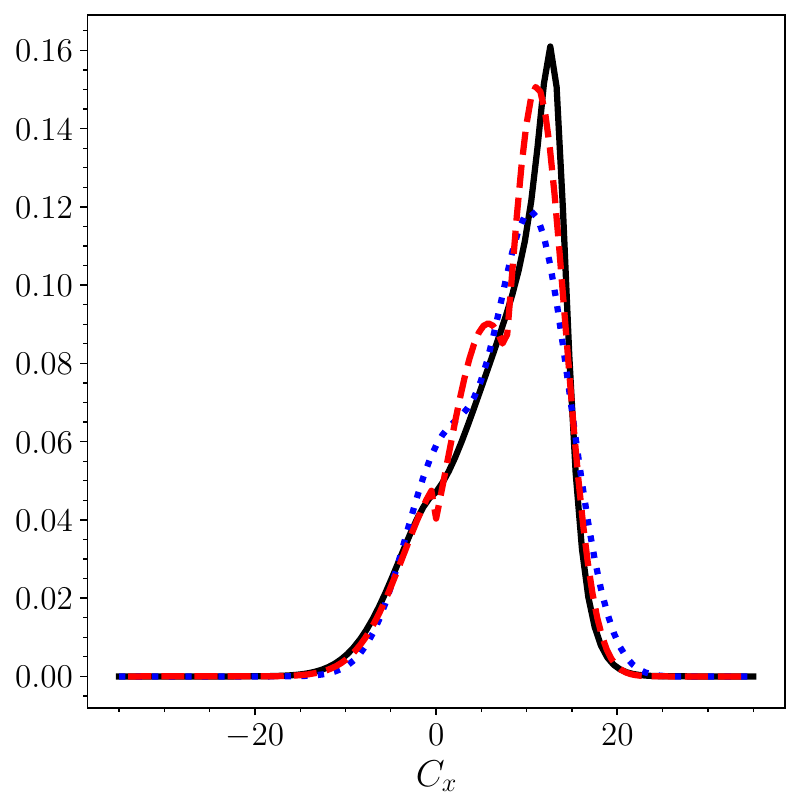}\label{fig:5c}}
    \subfloat[Mach 10, two \(x/\lambda_\infty\) upstream of center.]{\includegraphics[width=5.4cm]{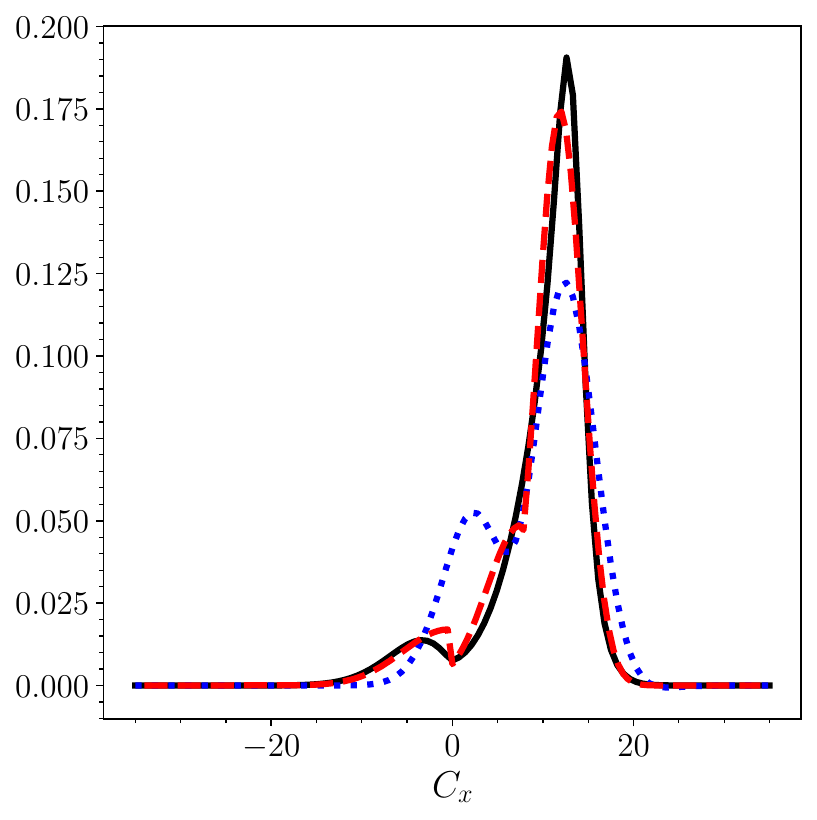}}
    \caption{Velocity distribution functions marginalized over y and z-axis using the MGME method and solutions obtained from the Boltzmann equation and the Chapman-Enskog solution of the Krook equation.}
    \label{fig:rom dist}
\end{figure}

In Fig.~\ref{fig:rom dist}, velocity distributions at various points in the shock marginalized over the y and z velocities is plotted. Distributions obtained using the MGME method, BGK model, and the Chapman-Enskog (C-E) expansion solution of the BGK model are compared with each other. The Chapman-Enskog expansion \cite{vincenti1965introduction} expresses deviations from equilibrium as a perturbative series in powers of Knudsen number where the zeroth, first, second power represents Euler, Navier-Stokes, and Burnett equations respectively. In this case, the perturbed equilibrium distribution is given by $f = f_0\left(1 + \phi\right)$ where $\phi$ is a function of the heat flux and shear stress \cite{vincenti1965introduction}. The heat flux and shear stress require temperature and velocity first order derivatives, which are taken from the BGK model solution. These derivatives can also be taken from solutions of the Navier-Stokes equations but will result in a final distribution that is less accurate. A different perturbation value is applied to the equilibrium distribution, $f_0$, at each spatial location. Solutions obtained for a free stream Mach number of 10 is presented to illustrate strong non-equilibrium. Since the C-E solution is valid for small deviations from equilibrium, a strong non-equilibrium case presents a significant challenge to the C-E solution. 

In the leftmost sub-figure of Fig.~\ref{fig:rom dist}, there is a bi-modality at the center of the shock that the C-E solution fails to capture, meanwhile, the three-grouped MGME method produces a bimodal solution but does not match the shape of the distribution. Two potential solutions to this issue exist: first, either more groups in velocity space are required, or second, the group bounds need to move such that a more favorable group distribution function can be solved for. Even so, the use of three groups is a significant improvement to the modelling of non-equilibrium behavior compared to the C-E expansion solution. The other two figures presented show the distribution one and two mean free paths upstream of the shock center. These distributions show excellent agreement with the underlying distribution and a significant improvement to the C-E solution. It can be seen that the choice of group bounds has a significant impact of how well the underlying BGK distribution can be modelled by group distributions. In these cases, the MGME model does a significantly better job than the C-E solution with appropriately chosen groups. At the boundaries where there is mostly equilibrium, the distribution can be accurately captured with one group. Because of this, groups bounds were chosen based on where non-equilibrium was expected to happen. Since a priori knowledge of the distribution is not always possible, different schemes to efficiently group the distribution will be explored in future works.

\subsubsection{Macroscopic properties}
\begin{figure}[ht]
    \centering
    \subfloat[\(M_\infty = 3\)]{\includegraphics[width=5.65cm]{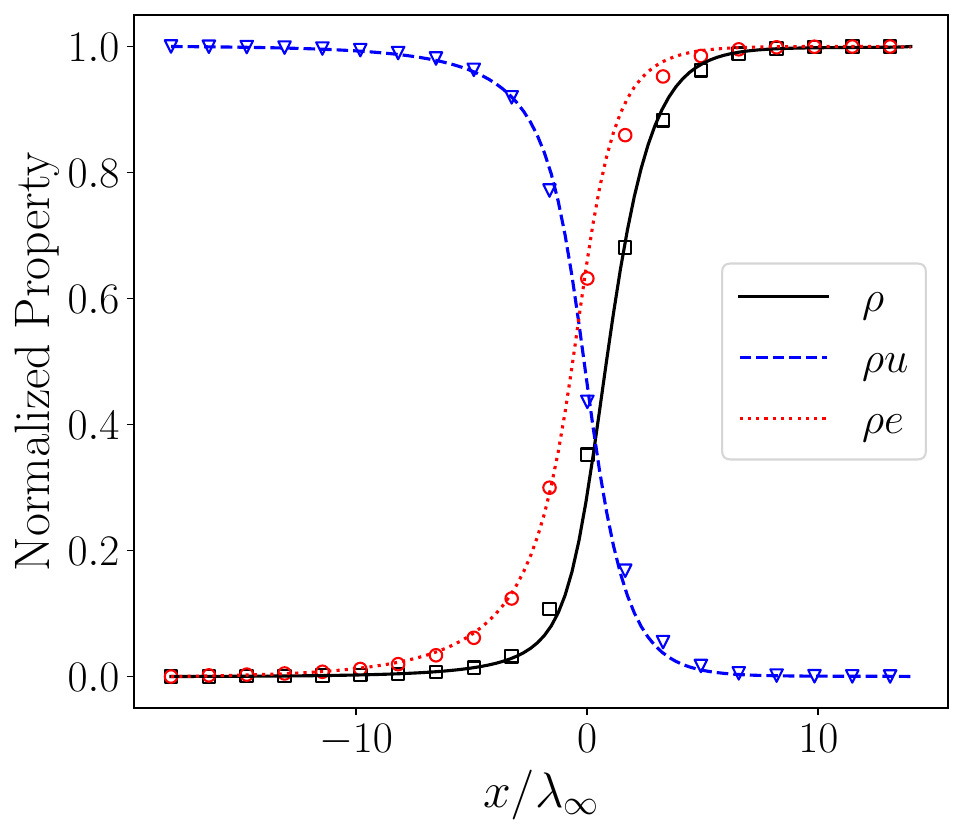}}
    \subfloat[\(M_\infty = 6\)]{\includegraphics[width=5.35cm]{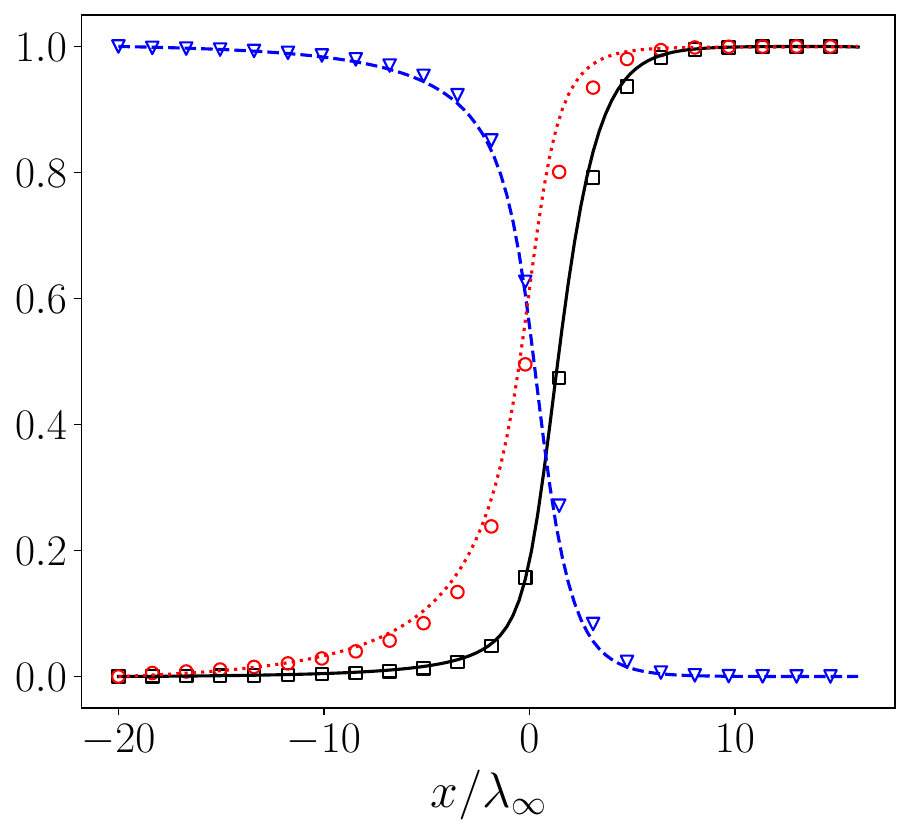}}
    \subfloat[\(M_\infty = 10\)]{\includegraphics[width=5.35cm]{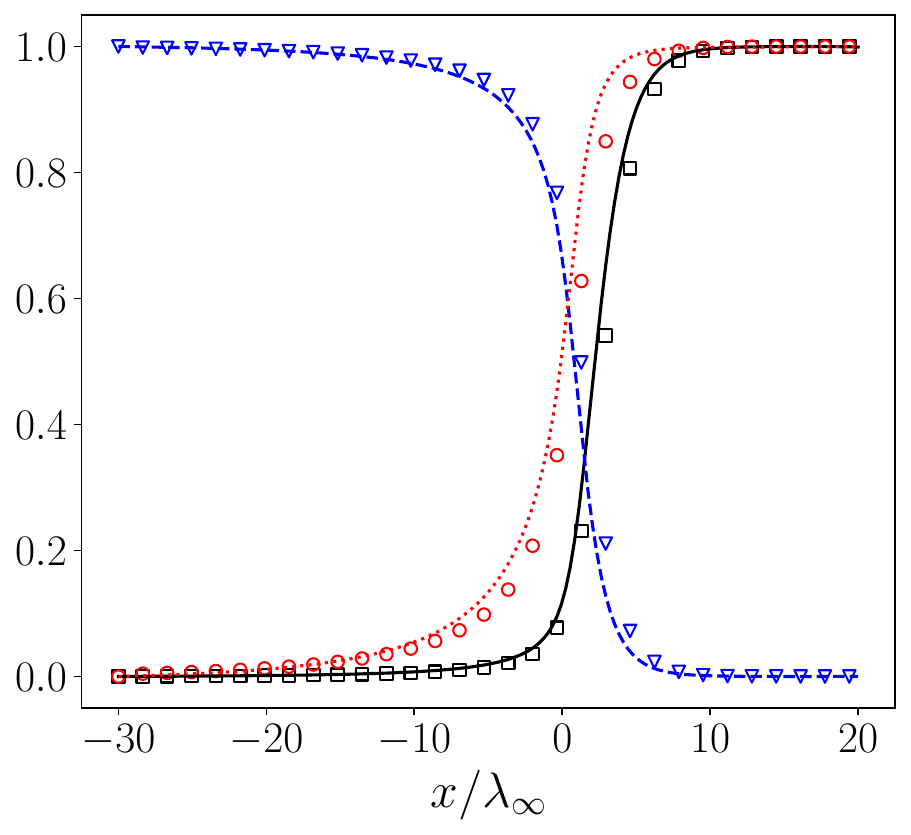}}
    \caption{Normalized macroscopic moments through the shock. The black, blue, and red lines represent the density, velocity, and temperature respectively. Lines denote solutions obtained from Boltzmann equation, and symbols denote solutions obtained from the MGME model.}
    \label{fig:shock_macro}
\end{figure}

\begin{figure}[ht]
    \centering
    \subfloat[\(M_\infty = 3\)]{\includegraphics[width=5.65cm]{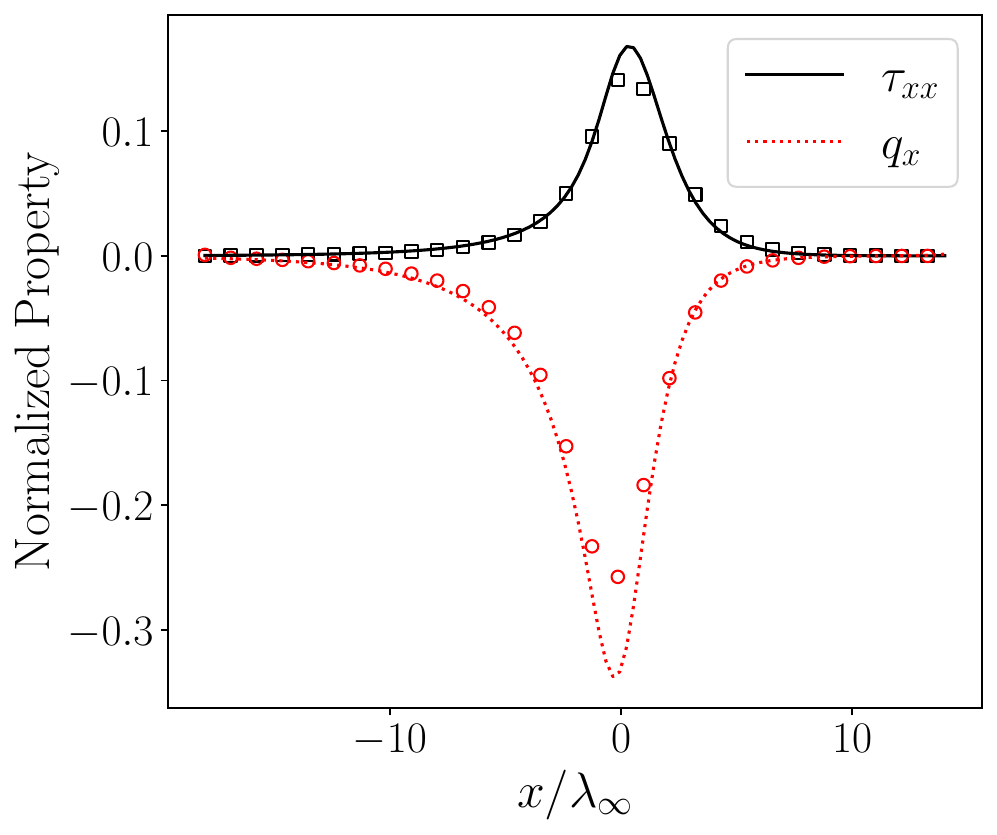}}
    \subfloat[\(M_\infty = 6\)]{\includegraphics[width=5.35cm]{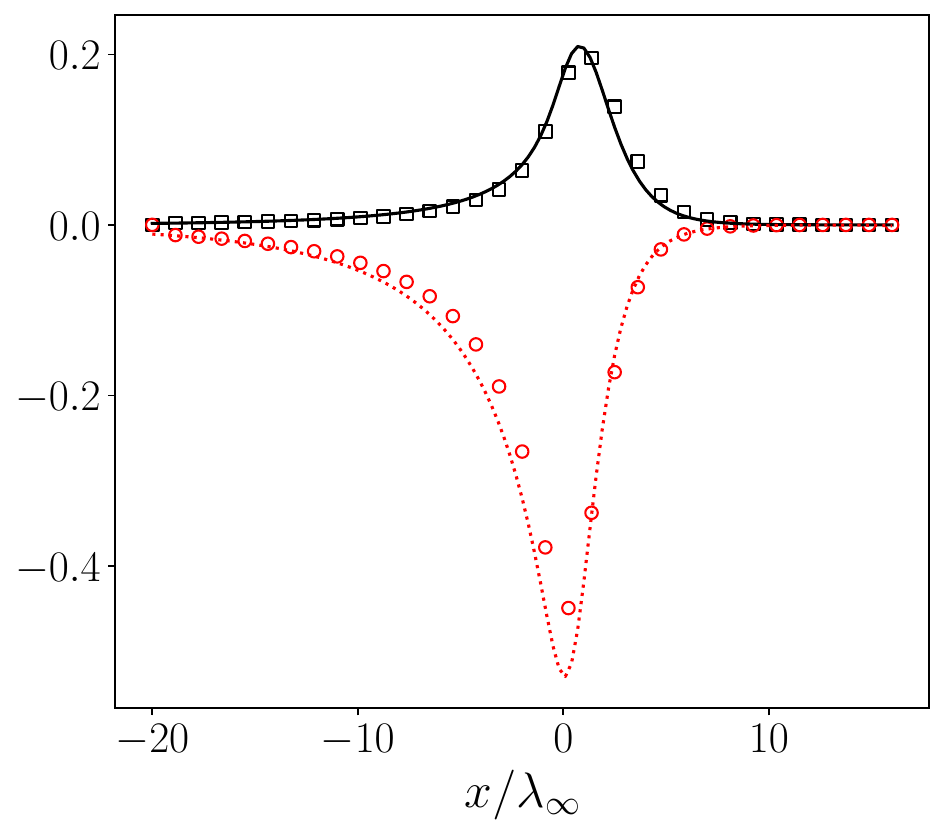}}
    \subfloat[\(M_\infty = 10\)]{\includegraphics[width=5.35cm]{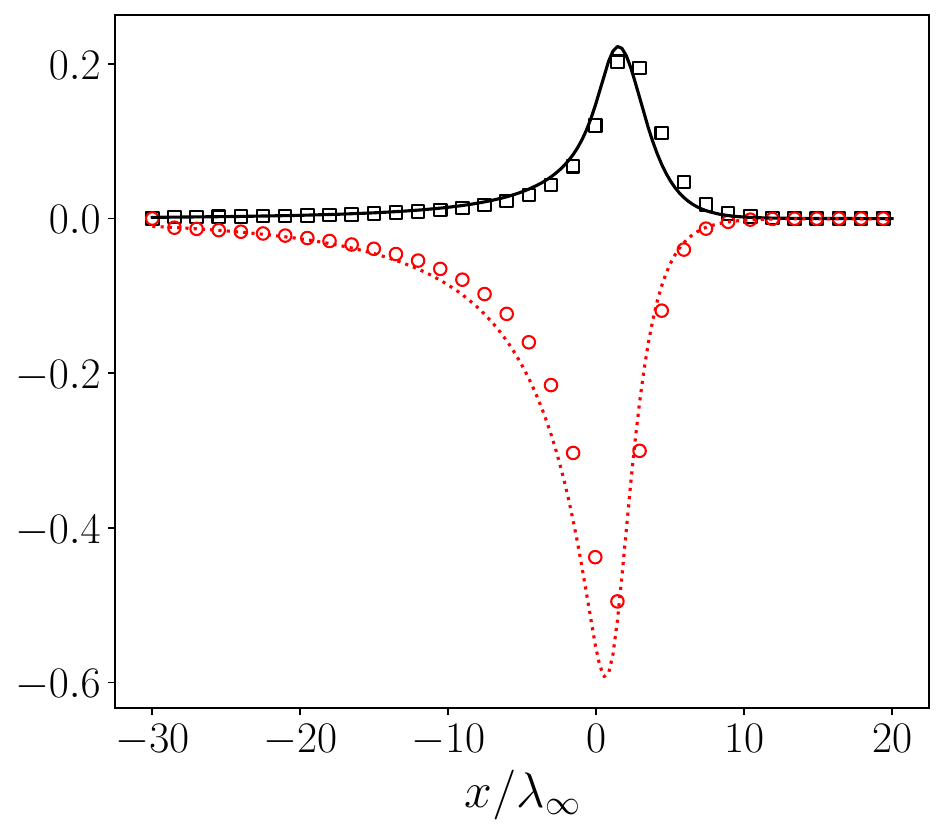}}
    \caption{Evolution of shear stress in the principal x-direction and heat flux in the x-direction with the varying pre-shock Mach number. (Line): BGK. (Symbols): MGME.}
    \label{fig:higher moments}
\end{figure}

The evolution of macroscopic properties through the shock is presented in this subsection. The density, momentum, and energy through the shock are presented in Fig.~\ref{fig:shock_macro}. All three quantities have been normalized to the range [0, 1] and are plotted with solutions obtained using the BGK model. There is excellent agreement between the MGME model and BGK model with more deviations as the moment order increases, with the most being seen in the second moment, $\rho e$. When there is very little or no non-equilibrium in the distribution, notably near the pre-shock and post-shock conditions, there is perfect agreement between the two models. The evolution of higher order moments, the normal shear stress and heat flux, are shown in Fig.~\ref{fig:higher moments}. The shear stress is normalized by the post-shock pressure, $P_2$, and the heat flux is normalized by $P_2 c_2/c_\text{ref}$ where $c_2$ is the post-shock velocity. The agreement between the MGME method and the BGK model is also good but has slightly larger discrepancies when compared with the macroscopic moments as these higher-order moments are more sensitive to errors in larger velocities. Overall, there is good agreement in the shear stress and heat flux although the maximum heat flux predicted by the MGME model is always less than that predicted by the BGK model. As with the macroscopic moments, there is perfect agreement in the pre-shock and post-shock equilibrium distributions. The ability to capture non-equilibrium behavior using only three velocity groups is remarkable since it results in a significant reduction in the velocity space order, from \(O(10^1 - 10^3)\) grid points in each direction to \(O(10^0 - 10^1)\) groups. Additional groups in the region of non-equilibrium would improve the agreement between the two models.

To quantify how well the MGME model captures each moment, the error compared to the BGK-Boltzmann equation, measured using the $L_2$ norm of the error, for density, velocity, temperature, shear stress, and heat flux is given by Tab.~\ref{tab:l2 norm macro} respectively for the range of considered Mach numbers. The relative error norm is shown in Tab.~\ref{tab:relative error macro} and is calculated for the center of the shock. The center of the shock is defined as the x-location where the scaled density is equal to 0.5. As the free stream Mach number is increased, the relative error increases for the macroscopic moments but shows no trend with the shear stress and heat flux. Depending on the Mach number, the location of the maximum shear stress and heat flux predicted by the MGME model is slightly shifted compared to the BGK model. Since the relative error is taken at a fixed point, this shift sometimes results in a coincidentally better performance at the shock center. At higher Mach numbers, the pre-shock and post-shock distributions which are further apart in velocity space cannot be as accurately captured by a small number of groups, resulting in increasing errors. The absolute $L_2$ norm of the error provides an overall comparison of the error between different Mach numbers and moments. Lower order moments, such as the density and velocity show relatively smaller errors compared to higher order moments. In the Mach 3 case, the growth of the error norm as the moment order is increased is small, only jumping significantly when considering the heat flux. However, in the Mach 10 case, the error jumps over two orders of magnitude from the density to the heat flux. The errors for each moment also grow as the Mach number is increased, with bigger growth seen as the order of the moments increases. 

\begin{table}[ht]
    \centering
    \begin{tabular}{p{1.5cm}|c|c|c}
    \hline
    \multirow{1}{1.5cm}{Moment} & \multicolumn{1}{c|}{Mach 3} & \multicolumn{1}{c|}{Mach 6} & \multicolumn{1}{c}{Mach 10} \\
    \hline
        \(n\) & 0.0388 & 0.0780 & 0.1962\\
        \(u\) & 0.0641 & 0.2529 & 0.5209 \\
        \(T\) & 0.0693 & 0.0950 & 0.1400 \\
        \(\tau_{xx}\) & 0.1259 & 0.0317 & 0.0146 \\
        \(q_x\) & 0.1957 & 0.0021 & 0.1388 \\
    \end{tabular}
    \caption{Relative error for density, velocity, temperature, shear stress, and heat flux for considered Mach numbers at the center of the shock. Error given in percentage deviation from BGK model.}
    \label{tab:relative error macro}
\end{table}

\begin{table}[ht]
    \centering
    \begin{tabular}{p{1.5cm}|c|c|c}
    \hline
    \multirow{1}{1.5cm}{Moment} & \multicolumn{1}{c|}{Mach 3} & \multicolumn{1}{c|}{Mach 6} & \multicolumn{1}{c}{Mach 10} \\
    \hline
        \(n\) & \num{1.539e-3} & \num{1.598e-3} & \num{3.006e-3} \\
        \(u\) & \num{1.688e-3} & \num{3.998e-3} & \num{1.054e-2} \\
        \(T\) & \num{2.152e-3} & \num{1.491e-2} & \num{4.933e-2} \\
        \(\tau_{xx}\) & \num{2.586e-3} & \num{1.416e-2} & \num{5.934e-2} \\
        \(q_x\) & \num{1.271e-2} & \num{1.040e-1} & \num{5.230e-1} \\
    \end{tabular}
    \caption{\(L_2\) norm of the error for density, velocity, temperature, shear stress, and heat flux for considered Mach. Error is divided by \(N_x\).}
    \label{tab:l2 norm macro}
\end{table}

\begin{figure}[ht]
    \centering
    \subfloat[Mach 10 total entropy.]{\includegraphics[width=7.7cm]{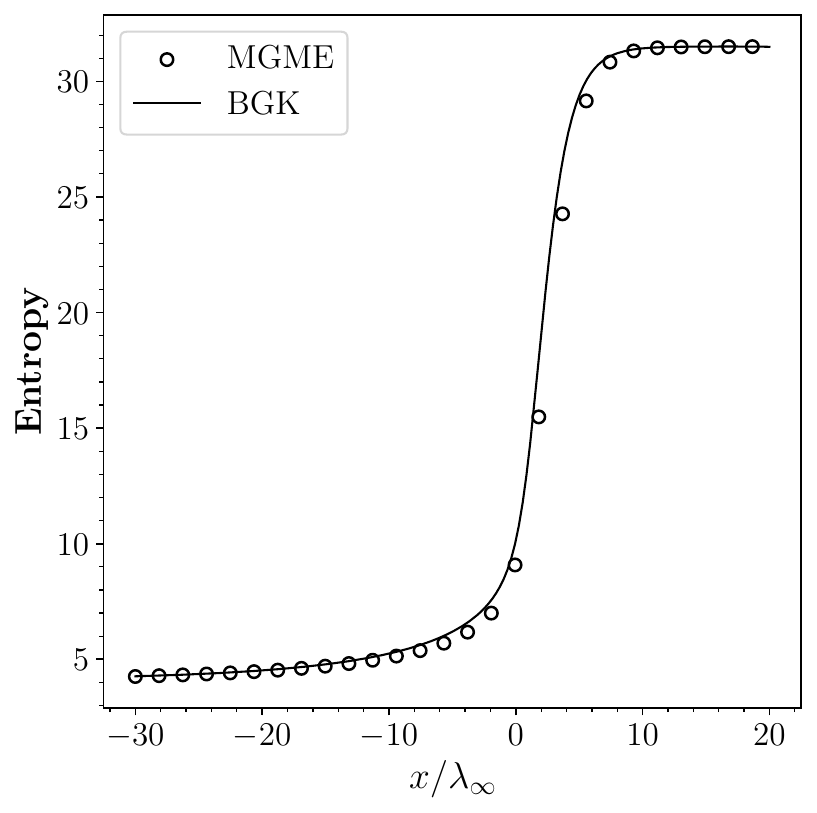}}
    \subfloat[Mach 10 group entropy.]{\includegraphics[width=7.3cm]{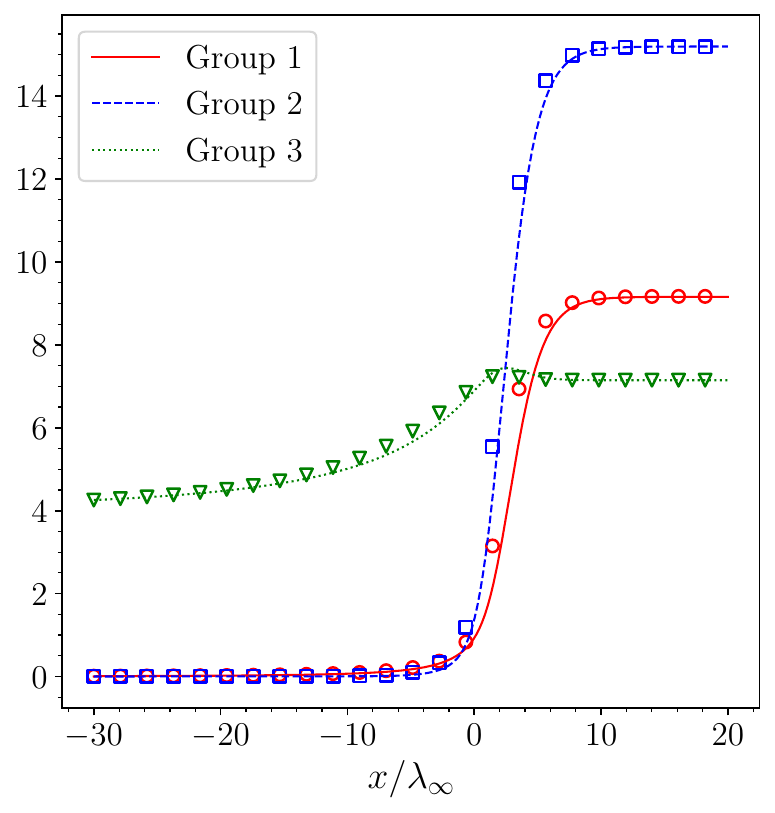}}
    \caption{Entropy evolution through shock wave. Grouping is the same as Fig.~\ref{fig:rom dist}. BGK solution is denoted by lines and MGME solutions is denoted with symbols.}
    \label{fig:entropy shock}
\end{figure}

Another important quantity to consider is the entropy evolution through the shock. In Fig.~\ref{fig:entropy shock}, the total entropy and group entropy for the Mach 10 case are plotted and compared with the BGK model. For the total entropy case, there is great agreement between the MGME model and BGK model. For the group entropy, there is also great agreement between the two models. The group entropy for the BGK case is calculated by applying Eq.~\ref{group entropy} using the same group bounds. At the pre-shock condition, group 3 contains the entirety of the pre-shock distribution. As a result, both groups 1 and 2 have zero entropy. As mass moves into these groups, the entropy increases with good agreement compared to the BGK model. In group 3, the entropy decreases slightly moving from pre-shock to post-shock. Therefore, while a group can have its entropy decrease, the overall entropy always increases due to the construction of the method. 

\subsubsection{Model validation using experimental data}
In addition to the macroscopic moments, derived shock properties are important in characterizing the behavior of a shock wave. These are the shock thickness and the density asymmetry, both of which have experimental data available.
The shock thickness changes with the free stream Mach number and proper modelling of the correct shock thickness requires non-equilibrium to be taken into account. Continuum approaches are unable to accurately capture the strong flow property discontinuities in the shock, resulting in under prediction of the shock thickness. The shock thickness is calculated using the maximum density gradient, which is defined as:
\begin{equation}
    \delta = \frac{\text{max}(\rho) - \text{min}(\rho)}{\text{max}\left(\partial \rho/\partial x\right)}
    \label{den gradient}
\end{equation}
\begin{figure}[ht]
    \centering
    \subfloat[Shock thickness]{\includegraphics[width=8.15cm]{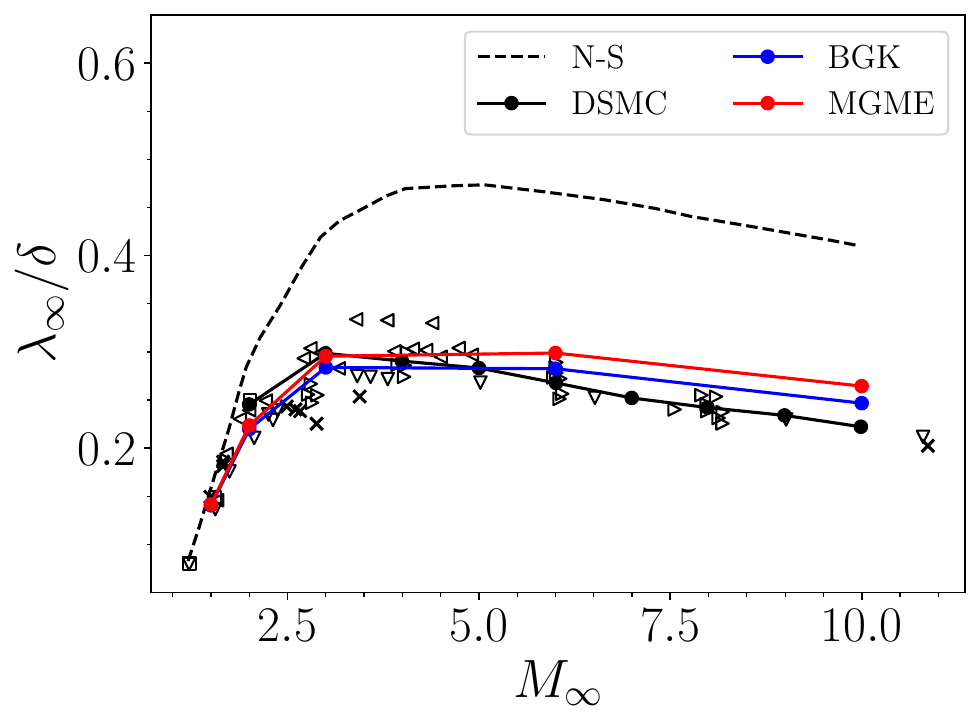}}
    \subfloat[Shock asymmetry]{\includegraphics[width=8.15cm]{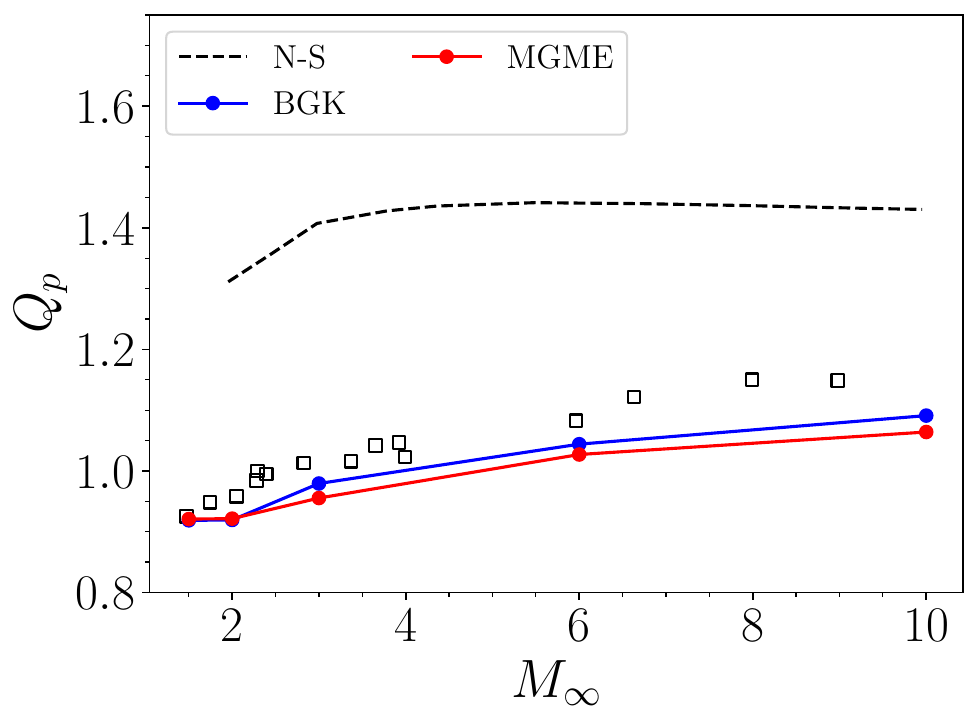}}
    \caption{Inverse shock thickness plotted against pre-shock Mach number. Experimental data is plotted using symbols. Shock asymmetry vs. pre-shock Mach number comparison between Navier-Stokes equations, DSMC, experimental data, BGK model, and the MGME equations.}
    \label{fig:shock thickness}
\end{figure}

Following Alsmeyer, $\delta$ is non-dimensionalized by the freestream mean free path, given by Eq.~\ref{ref mean free path}, and is plotted as the inverse shock thickness, $\lambda_\infty/\delta$. Fig.~\ref{fig:shock thickness} compares our results with DSMC results \cite{macart2022deep}, the Navier-Stokes equations \cite{macart2022deep}, and a collection of experimental data from Alsmeyer \cite{alsmeyer1976density}. The experimental data is plotted using symbols and represents experimental data taken from various sources and compiled by Alsmeyer. The shock thickness for the Mach 3, 6, and 10 cases are plotted along with additional runs done at Mach 1.5 and 2 to show the complete behavior of the shock thickness. For the additional runs, three groups are similarly used. The MGME model shows a marked improvement over the Navier-Stokes equations, which over-predict the inverse shock thickness at Mach numbers over around 1.5. The model performs excellently at Mach numbers below 3 and over-predicts slightly at higher Mach numbers. There is overall good agreement with the BGK model across all Mach numbers tested with the use of three groups, with potential to improve the solution with a different choice of group bounds.

The density asymmetry quotient is another derived quantity that assesses the overall shape of the density profile. This quantity is more sensitive to the detailed microscopic physics and is defined as:
\begin{equation}
    Q_p = \frac{\int_{-\infty}^0 \rho^* \,dx}{\int_0^\infty \left(1 - \rho^*\right) \,dx}
\end{equation}
\noindent where \(\rho^*\) is the normalized density centered at x = 0. The asymmetry quotient measures the skewness of the profile relative to its midpoint. A symmetrical profile would have \(Q_p = 1\). 
In Fig.~\ref{fig:shock thickness}, the density asymmetry quotient for the Navier-Stokes equations, experimental data \cite{alsmeyer1976density}, the target BGK model, and the MGME model is plotted. The Navier-Stokes predicts upstream-skewed shocks for all Mach numbers, while the experimental data predicts upstream-skewed shocks for Mach numbers greater than 2.5 and downstream-skewed shocks for Mach numbers smaller than 2.5. The BGK model performs decently well across all Mach numbers but tends to predict more downstream-skewed shocks compared to experimental data. The MGME model shows excellent agreement with the BGK model at low Mach numbers with some deviation at Mach 3 and above. Overall, the performance of the MGME method with three groups presents an significant reduction in computation while retaining the ability to capture many of the flow features, with an increase in accuracy expected as the number of groups increases.  

\section{Conclusions and Future Work}
This work presents a maximum entropy-based closure for a coarse-grained model of the Krook equation. The model divides the velocity space into groups, and group distribution functions are solved for using the maximum entropy principle within each group. Macroscopic moments, density, velocity, and temperature are used to solve for the distribution function in each group. This approach allows for the modeling of non-equilibrium effects while still dealing with bulk flow properties. The ability of the coarse-grained model to recreate the BGK model through the use of a small number of groups is shown through space-homogeneous and space in-homogeneous test cases. In the standing shock case, good agreement between the BGK model and the MGME equations is shown across macroscopic moments, higher-order moments, shock thickness, and shock profile asymmetry. Results from the MGME equations have also been compared with results obtained using the Navier-Stokes equations, experimental data, and DSMC data.

Future work will also focus on adapting this model for the full binary collision operator, implementing an upwind or higher-order flux scheme, and a faster implementation of the non-linear inversion. This would extend the model to cases where the BGK model is insufficient, provide a more accurate description of the gas, and increase computational speed. The extension of this method to two- and three-dimensional space would extend the use of this method and allow for more test cases from the literature to be attempted. The maximum entropy-based closure can also be trained using deep learning-based tools, either at the distribution function or macroscopic level.

\section{Acknowledgment}
I would like to thank Vegnesh Jayaraman and Dr. Yen Liu for the initial development of the code. This work is supported by the Vannevar Bush Faculty Fellowship OUSD(RE) Grant No: N00014-21-1-295 with Prof. Marco Panesi as the Principal Investigator.
\appendix
\section{Appendix }
\label{app1}

\subsection{Definition of particular integrals}
\begin{equation}
    \mathfrak{I}^{0}_i [c^k_i, c^{k+1}_i] = \sqrt{\frac{\pi}{4\beta}}\left[\erf\left(\sqrt{\beta}\; \left(c^{k+1}_i - w_i\right)\right) - \erf\left(\sqrt{\beta} \; \left(c^{k}_i - w_i)\right) \right)\right]
\end{equation}
\begin{equation}
    \mathfrak{I}^{1}_i[c^k_i, c^{k+1}_i]=  \frac{1}{2\beta} \left[\exp\left(-\beta \ \left(c^k_i - w_i\right)^2\right) - \exp\left(-\beta \; \left(c^{k+1}_i - w_i\right)^2 \right)\right]
\end{equation}
\begin{equation}
    \mathfrak{I}^{2}_i[c^k_i, c^{k+1}_i] = \frac{1}{2\beta}\left[\left(c^{k+1}_i - w_i\right)\, \exp\left(-\beta \,\left(c^{k+1}_i - w_i\right)^2\right) - \left(c^k_i - w_i\right)\, \exp\left(\sqrt{\beta} \,\left(c^k_i - w_i\right)^2\right)+\mathfrak{I}^{0}_i[c^k_i, c^{k+1}_i]\right]
\end{equation}
\begin{equation}
    \mathfrak{I}^{3}_i[c^k_i, c^{k+1}_i] = \frac{1}{\beta}\, \mathfrak{I}^{1}_i[c^k_i, c^{k+1}_i]  +\frac{1}{2 \beta}\left(\left(c^k_i - w_i\right)^2 \exp\left({-\beta\left(c^k_i - w_i\right)^2}\right) - \left(c^{k+1}_i - w_i\right)^2 \exp\left(-\beta \left(c^{k+1}_i - w_i\right)^2\right)\right)
\end{equation}

\subsection{Flux function}

\[
\int_{c_{x_a}}^{c_{x_b}} \int_{c_{y_a}}^{c_{y_b}} \int_{c_{z_a}}^{c_{z_b}} (\mathbf{c} \cdot \mathbf{c}) \mathbf{c} f(\mathbf{c}) \, dc_x \, dc_y \, dc_z
\]
where \( f(\mathbf{c}) \) is a function of the vector \( \mathbf{c} \), and \( c_{x_a}, c_{x_b}, c_{y_a}, c_{y_b}, c_{z_a}, c_{z_b} \) are the bounds of integration for each component of \( \mathbf{c} \).

\section{Case 1: \( f(\mathbf{c}) = A \exp(-\beta (\mathbf{c} \cdot \mathbf{c})) \)}

The integral becomes
\[
\int_{c_{x_a}}^{c_{x_b}} \int_{c_{y_a}}^{c_{y_b}} \int_{c_{z_a}}^{c_{z_b}} (\mathbf{c} \cdot \mathbf{c}) \mathbf{c} A \exp(-\beta (\mathbf{c} \cdot \mathbf{c})) \, dc_x \, dc_y \, dc_z
\]
The Wolfram Language could not provide a symbolic solution for this integral, suggesting it might not have a closed-form expression.

\section{Case 2: Expanded Integral}

We consider the expanded integrand \( (\mathbf{v} \cdot \mathbf{v}) + 2 \mathbf{c} \cdot \mathbf{w} - w^2 \) and \( \mathbf{v} + \mathbf{w} \). The integrand becomes
\[
(\mathbf{v} \cdot \mathbf{v}) \mathbf{v} + (\mathbf{v} \cdot \mathbf{v}) \mathbf{w} + 2 \mathbf{c} \cdot \mathbf{w} \mathbf{v} + 2 \mathbf{c} \cdot \mathbf{w} \mathbf{w} - w^2 \mathbf{v} - w^2 \mathbf{w}
\]
This integral can be split into six separate integrals:
\begin{enumerate}
    \item \( \int (\mathbf{v} \cdot \mathbf{v}) \mathbf{v} f \, d\mathbf{c} \)
    \item \( \int (\mathbf{v} \cdot \mathbf{v}) \mathbf{w} f \, d\mathbf{c} \)
    \item \( \int 2 \mathbf{c} \cdot \mathbf{w} \mathbf{v} f \, d\mathbf{c} \)
    \item \( \int 2 \mathbf{c} \cdot \mathbf{w} \mathbf{w} f \, d\mathbf{c} \)
    \item \( \int -w^2 \mathbf{v} f \, d\mathbf{c} \)
    \item \( \int -w^2 \mathbf{w} f \, d\mathbf{c} \)
\end{enumerate}

\subsection{Problem Statement}

Given \( f(v) = A \exp(-\beta v^2) \), where \( A \) and \( \beta \) are constants, and \( v^2 = v_x^2 + v_y^2 + v_z^2 \), the objective is to solve the following integral:

\[
\int_{v_{a_x}}^{v_{b_x}} \int_{v_{a_y}}^{v_{b_y}} \int_{v_{a_z}}^{v_{b_z}} v^2 \mathbf{v} f(v) \, dv_x \, dv_y \, dv_z
\]

\section{Integral Decomposition}

The integral can be decomposed into its Cartesian components:

\begin{align*}
&\int_{v_{a_x}}^{v_{b_x}} \int_{v_{a_y}}^{v_{b_y}} \int_{v_{a_z}}^{v_{b_z}} A v_x^3 \exp(-\beta v^2) \, dv_x \, dv_y \, dv_z, \\
&\int_{v_{a_x}}^{v_{b_x}} \int_{v_{a_y}}^{v_{b_y}} \int_{v_{a_z}}^{v_{b_z}} A v_y^3 \exp(-\beta v^2) \, dv_x \, dv_y \, dv_z, \\
&\int_{v_{a_x}}^{v_{b_x}} \int_{v_{a_y}}^{v_{b_y}} \int_{v_{a_z}}^{v_{b_z}} A v_z^3 \exp(-\beta v^2) \, dv_x \, dv_y \, dv_z
\end{align*}

\section{Analytical Solution for the First Integral}

The first integral can be solved analytically to yield:

\[
\frac{A \pi}{8 \beta^3} \left( e^{-\beta v_{a_x}^2} (1 + \beta v_{a_x}^2) - e^{-\beta v_{b_x}^2} (1 + \beta v_{b_x}^2) \right) \left( \text{Erf}[\sqrt{\beta} v_{a_y}] - \text{Erf}[\sqrt{\beta} v_{b_y}] \right) \left( \text{Erf}[\sqrt{\beta} v_{a_z}] - \text{Erf}[\sqrt{\beta} v_{b_z}] \right)
\]

\section{Substitution of \( I_3 \)}

We define \( I_3 \) as:

\[
I_3 = \frac{1}{2 \beta^2} \left( e^{-\beta v_a^2} - e^{-\beta v_b^2} \right) + \frac{1}{2 \beta} \left( v_a^2 e^{-\beta v_a^2} - v_b^2 e^{-\beta v_b^2} \right)
\]

Substituting \( I_3 \) into the original expression, we get:

\[
\frac{A \pi I_3}{4 \beta} \left( \text{Erf}[\sqrt{\beta} v_{a_y}] - \text{Erf}[\sqrt{\beta} v_{b_y}] \right) \left( \text{Erf}[\sqrt{\beta} v_{a_z}] - \text{Erf}[\sqrt{\beta} v_{b_z}] \right)
\]

\begin{equation}
    \mathfrak{I}^{0}[v]= \sqrt{\frac{\pi}{4\beta}}\left[\text{Erf}\left(\sqrt{\beta}\; v_f\right) - \text{Erf}\left(\sqrt{\beta} \; v_i \right)\right]
\end{equation}

\[
\frac{A \pi I_3 (v_x)}{4 \beta} \; \left(\frac{4\beta}{\pi}\right) \mathfrak{I}^0_y \mathfrak{I}^0_z
\]

$$
\int_{v_{a_x}}^{v_{b_x}} \int_{v_{a_y}}^{v_{b_y}} \int_{v_{a_z}}^{v_{b_z}} A v_x^3 \exp(-\beta v^2) \, dv_x \, dv_y \, dv_z = \frac{I_3 (v_x)}{I_0 (v_x)}
$$

The total integral is:

$$
\int_{v_{a_x}}^{v_{b_x}} \int_{v_{a_y}}^{v_{b_y}} \int_{v_{a_z}}^{v_{b_z}}  ({\bf v} \cdot {\bf v}) \; {\bf v} \; \exp(-\beta v^2) \, dv_x \, dv_y \, dv_z = \frac{I_3 (v_x)}{I_0 (v_x)} \; {\bf i} \;\;  + 
\frac{I_3 (v_y)}{I_0 (v_y)} \; {\bf j} \; + \;  \frac{I_3 (v_z)}{I_0 (v_z)}  \; {\bf k} 
$$

Second integral 

$$
\int_{v_{a_x}}^{v_{b_x}} \int_{v_{a_y}}^{v_{b_y}} \int_{v_{a_z}}^{v_{b_z}}  ({\bf v} \cdot {\bf v}) \; {\bf w} \; \exp(-\beta v^2) \, dv_x \, dv_y \, dv_z = \; {\bf w}  \int_{v_{a_x}}^{v_{b_x}} \int_{v_{a_y}}^{v_{b_y}} \int_{v_{a_z}}^{v_{b_z}}  ({\bf v} \cdot {\bf v}) \; \exp(-\beta v^2) \, dv_x \, dv_y \, dv_z \\
 = \; \langle {\bf v} \cdot {\bf v}   \rangle {\bf w} 
$$

Third integral 

$$
\int_{v_{a_x}}^{v_{b_x}} \int_{v_{a_y}}^{v_{b_y}} \int_{v_{a_z}}^{v_{b_z}} 2 ({\bf c} \cdot {\bf w}) \; {\bf v} \; \exp(-\beta v^2) \, dv_x \, dv_y \, dv_z = \; 2 \int_{v_{a_x}}^{v_{b_x}} \int_{v_{a_y}}^{v_{b_y}} \int_{v_{a_z}}^{v_{b_z}}  [({\bf v} + {\bf w})  \cdot {\bf w}] \; \exp(-\beta v^2) \, dv_x \, dv_y \, dv_z 
$$

$$
= \; 2 \int_{v_{a_x}}^{v_{b_x}} \int_{v_{a_y}}^{v_{b_y}} \int_{v_{a_z}}^{v_{b_z}}[({\bf v} \cdot {\bf w}) + w^2  ]  {\bf v} \; \exp(-\beta v^2) \, dv_x \, dv_y \, dv_z  
$$

$$
= \; 2 \int_{v_{a_x}}^{v_{b_x}} \int_{v_{a_y}}^{v_{b_y}} \int_{v_{a_z}}^{v_{b_z}}({\bf v} \cdot {\bf w})  {\bf v} \; \exp(-\beta v^2) \, dv_x \, dv_y \, dv_z  + 2 w^2 {\bf u}
$$

$$
= \; 2 \int_{v_{a_x}}^{v_{b_x}} \int_{v_{a_y}}^{v_{b_y}} \int_{v_{a_z}}^{v_{b_z}} v_x ({\bf v} \cdot {\bf w})   \; \exp(-\beta v^2) \, dv_x \, dv_y \, dv_z   $$

$$
=  \; 2 \int_{v_{a_x}}^{v_{b_x}} \int_{v_{a_y}}^{v_{b_y}} \int_{v_{a_z}}^{v_{b_z}} v_x (v_x w_x + v_y w_y + v_z w_z)   \; \exp(-\beta v^2) \, dv_x \, dv_y \, dv_z
$$

$$
=  \; 2 \int_{v_{a_x}}^{v_{b_x}} \int_{v_{a_y}}^{v_{b_y}} \int_{v_{a_z}}^{v_{b_z}} v^2_x w_x    \; \exp(-\beta v^2) \, dv_x \, dv_y \, dv_z + \; 2 \int_{v_{a_x}}^{v_{b_x}} \int_{v_{a_y}}^{v_{b_y}} \int_{v_{a_z}}^{v_{b_z}} v_x v_y w_y    \; \exp(-\beta v^2) \, dv_x \, dv_y \, dv_z + \ldots
$$

$$
=  \; 2 \left[ w_x \mathfrak{I}^{2}(v_x) + w_y \mathfrak{I}^1_x \,   \mathfrak{I}^1_y  \, \mathfrak{I}^0_z +  w_z \mathfrak{I}^1_x \,   \mathfrak{I}^1_z  \, \mathfrak{I}^0_y \right] \; {\bf i}
$$
$$
\; + \; 2 \, \left[ w_x \mathfrak{I}^1_x \,   \mathfrak{I}^{1} (v_y) \, \mathfrak{I}^0_z + w_y \mathfrak{I}^{2}(v_y)  + w_z \mathfrak{I}^1_y \,   \mathfrak{I}^1_z  \, \mathfrak{I}^0_x \right]  \; {\bf j}
$$
$$
\; + \; 2 \, \left[ w_x \mathfrak{I}^1_x \,   \mathfrak{I}^{1} (v_z) \, \mathfrak{I}^0_y + w_y \mathfrak{I}^1_y \,   \mathfrak{I}^{1} (v_z) \, \mathfrak{I}^0_x  \, + w_z  \mathfrak{I}^{2}(v_z)  \right]  \; {\bf k}
$$






\bibliographystyle{elsarticle-num}
\bibliography{sample}

\end{document}